%% file: 01_main.tex
\documentclass{article}
\usepackage{arxiv}

\usepackage[utf8]{inputenc}
\usepackage[T1]{fontenc}    
\usepackage{amsmath, amssymb, amsfonts, graphicx, multirow, color, textcomp, subfig, xcolor, makecell, enumitem, totcount}
\usepackage{url}
\usepackage[colorlinks = true, citecolor = burntorange, urlcolor = airforceblue, linkcolor = airforceblue]{hyperref}
\usepackage{cellspace}
\usepackage{tabularx}

\usepackage{booktabs}
\usepackage{natbib}
\usepackage[normalem]{ulem}

\definecolor{burntorange}{rgb}{0.8, 0.33, 0.0}
\definecolor{airforceblue}{rgb}{0.36, 0.54, 0.66}

\renewcommand{\headeright}{2nd Order Solutions}
\renewcommand{\undertitle}{2nd Order Solutions\thanks{The authors would like to thank 2nd Order Solutions for their continued support through the inception and realization of this project. 2nd Order Solutions is a boutique credit risk consultancy firm that advises clients ranging from top 10 banks to fintechs throughout the world. }}
\renewcommand{\shorttitle}{A comparison of modeling preprocessing techniques}

\hypersetup{
pdftitle = {A comparison of modeling preprocessing techniques},
pdfauthor = {Tosan ~Johnson, Alice J. ~Liu, Syed ~Raza},
pdfkeywords = {feature selection, categorical encoding, null imputation, preprocessing, xgboost},
}

\newcounter{findno}

\regtotcounter{findno}

\newcounter{obsno}
\newcommand{\obs}[1]{\refstepcounter{obsno}\label{#1}}

\regtotcounter{obsno}
\newcommand*{\obscount}{\total{obsno}}

\graphicspath{ {./images/} }

\title{A comparison of modeling preprocessing techniques}

\begin{document}

\makeatletter
\newcommand{\linebreakand}{%
  \end{@IEEEauthorhalign}
  \hfill\mbox{}
  \mbox{}\hfill\begin{@IEEEauthorhalign}
}
\makeatother

\author{\\
\textbf{Tosan Johnson}\\
  \textit{Senior Business Analyst}\\
    \textit{2\textsuperscript{nd} Order Solutions}\\
    tosan.johnson@2os.com
  \and \\
  \textbf{Alice J. Liu, Ph.D.}\\
  \textit{Senior Data Scientist}\\
    \textit{2\textsuperscript{nd} Order Solutions}\\
    alice.liu@2os.com
  \and \\
  \textbf{Syed Raza, Ph.D.}\\
  \textit{Data Science Manager}\\
    \textit{2\textsuperscript{nd} Order Solutions}\\
    syed.raza@2os.com
  \and \\
  \textbf{Aaron McGuire}\\
  \textit{Director of Data Science}\\
    \textit{2\textsuperscript{nd} Order Solutions}\\
    aaron.mcguire@2os.com
}

\date{\today}

\maketitle
\begin{abstract}
This paper compares the performance of various data processing methods in terms of predictive performance for structured data. This paper also seeks to identify and recommend preprocessing methodologies for tree-based binary classification models, with a focus on eXtreme Gradient Boosting (XGBoost) models. Three data sets of various structures, interactions, and complexity were constructed, which were supplemented by a real-world data set from the Lending Club. We compare several methods for feature selection, categorical handling, and null imputation. Performance is assessed using relative comparisons among the chosen methodologies, including model prediction variability. This paper is presented by the three groups of preprocessing methodologies, with each section consisting of generalized observations. Each observation is accompanied by a recommendation of one or more preferred methodologies.

\quad Among feature selection methods, permutation-based feature importance, regularization, and XGBoost's feature importance by weight are not recommended. The correlation coefficient reduction also shows inferior performance. Instead, XGBoost importance by gain shows the most consistency and highest caliber of performance. Categorical featuring encoding methods show greater discrimination in performance among data set structures. While there was no universal ``best'' method, frequency encoding showed the greatest performance for the most complex data sets (Lending Club), but had the poorest performance for all synthetic (i.e., simpler) data sets. Finally, missing indicator imputation dominated in terms of performance among imputation methods, whereas tree imputation showed extremely poor and highly variable model performance.
\end{abstract}
\newpage
\tableofcontents

\newpage
\section{Introduction}\label{sec:introduction}
\input{02_introduction}

\section{Findings}
\subsection{Feature Selection}\label{sec:feature}
\input{03a_feature_selection}

\subsection{Categorical Handling}\label{sec:categorical}
\input{03b_categorical}

\subsection{Null Imputations}\label{sec:null}
\input{03c_null_imputation}

\section{Limitations}\label{sec:limitations}
\input{04_limitations}

\section{Conclusions}\label{sec:conclusion}
\input{05_conclusion}

\bibliography{refs}

\newpage
\appendix
\input{06_appendix}

\end{document}

%% file: 02_introduction.tex
Over the past decade, large banks and fintechs have become increasingly reliant on data analytics and machine learning to make informed decisions regarding safe and effective money lending to consumers. With companies pouring millions of dollars into data collection and infrastructure resources, a record volume of data is flowing into models and decision engines that have a direct impact on revenue and losses. To ensure the quality and utility of this data, data scientists have developed hundreds of techniques to apply to their input data sets, squeezing out maximum performance of their models. These techniques are commonly referred to as ``preprocessing.''

\quad Our goal in this paper is to identify and explain behaviors observed for a variety of preprocessing methodologies across three distinct categories: feature selection, categorical encoding, and null imputation. We explore the empirical behavior of popular preprocessing methods to provide a deeper understanding of the selected methodologies.

\quad While the tangible results from this paper are the empirical observations and recommendations made for feature selection, categorical encoding, and null imputation methods, the work completed here lays the forward-looking foundation to conduct future investigation in similar or adjacent areas.\footnote{This study is the proof-of-concept of the innovative research that is currently in progress at 2nd Order Solutions. We work to ensure our clients receive expert recommendations and knowledge, backed by empirical and academic research.} We tested the following preprocessing methodologies. Please refer to Appendix \ref{app:methods} for more details on each of the methods used. 

\begin{table}[h!]
    \begin{center}
        \begin{tabular}{c c c} 
             \toprule
              \bf{Feature Selection Methods} & \bf{Categorical Encoding Methods} & \bf{Null Imputation Methods} \\ [0.5ex] 
             \midrule
             Pearson correlation   & One-hot encoding  & Mean imputation  \\
             Spearman’s rank correlation  & Helmert coding   & Median imputation  \\
              XGB importance (weight and gain) & Frequency encoding  & Missing indicator imputation  \\
              Regularization &  Binary encoding  & Decile imputation  \\
              Permutation-based importance &    & Clustering imputation  \\
              RFE  &    & Decision tree imputation  \\
             \bottomrule
        \end{tabular} \vspace{2mm}
    \end{center}
    \caption{Preprocessing methodologies used in this study}
    \label{tab:methods}
\end{table}

\quad We tested these methodologies on four different types of data sets; three synthetic data sets generated with varying levels of complexity, and a real-world data set from Lending Club. Refer to Appendix \ref{app:data} for details and methodology behind these data sets.    

\begin{table}[h!]
    \begin{center}
        \begin{tabular}{c c c c c} 
             \toprule
             \textbf{Data Set} & \textbf{Data Set Type} \\ [0.5ex] 
             \midrule
             Linear & Synthetic  \\ 
             GAM Global & Synthetic  \\
             Jumpy GAM Local & Synthetic  \\
             Lending Club & Real-world  \\
             \bottomrule
        \end{tabular} \vspace{2mm}
    \end{center}
    \caption{Types of data sets used in this study}
    \label{tab:data}
\end{table}

\quad We introduce and discuss \obscount{} observations and their associated implications in practical use. We list the main findings below, grouped by the preprocessing methodology.

\begin{description}[style = nextline]
    \item[Feature Selection]
    \nameref{obs:feat1}. \\
    \\
    \nameref{obs:feat2}. \\
    \\
    \nameref{obs:feat3}. \\
    \\
    \nameref{obs:feat4}. \\
    \\
    \nameref{obs:feat5}. \\
    
    \item[Categorical Encoding]
        \nameref{obs:cat1}. \\
        \\
        \nameref{obs:cat2}. \\
    
    \item[Null Imputation]
        \nameref{obs:null1}. \\
        \\
        \nameref{obs:null2}. \\
        \\
        \nameref{obs:null3}.
        
\end{description}

%% file: 03a_feature_selection.tex
\begin{description}[style = nextline]

    \item[Observation \ref{featselobs1}: The choice of feature selection method is trivial for simple data structures. \label{obs:feat1}] \obs{featselobs1}
    The results from the three synthetic data sets suggest that the choice of feature selection method does not necessarily matter in terms of AUC performance. In general, all feature selection methods more or less choose the same subset of features, resulting in nearly identical performance across the compared methods (see Figure \ref{fig:featsel_linear2}).
    \begin{figure}[h!]
        \centering
        \includegraphics[scale = 0.25]{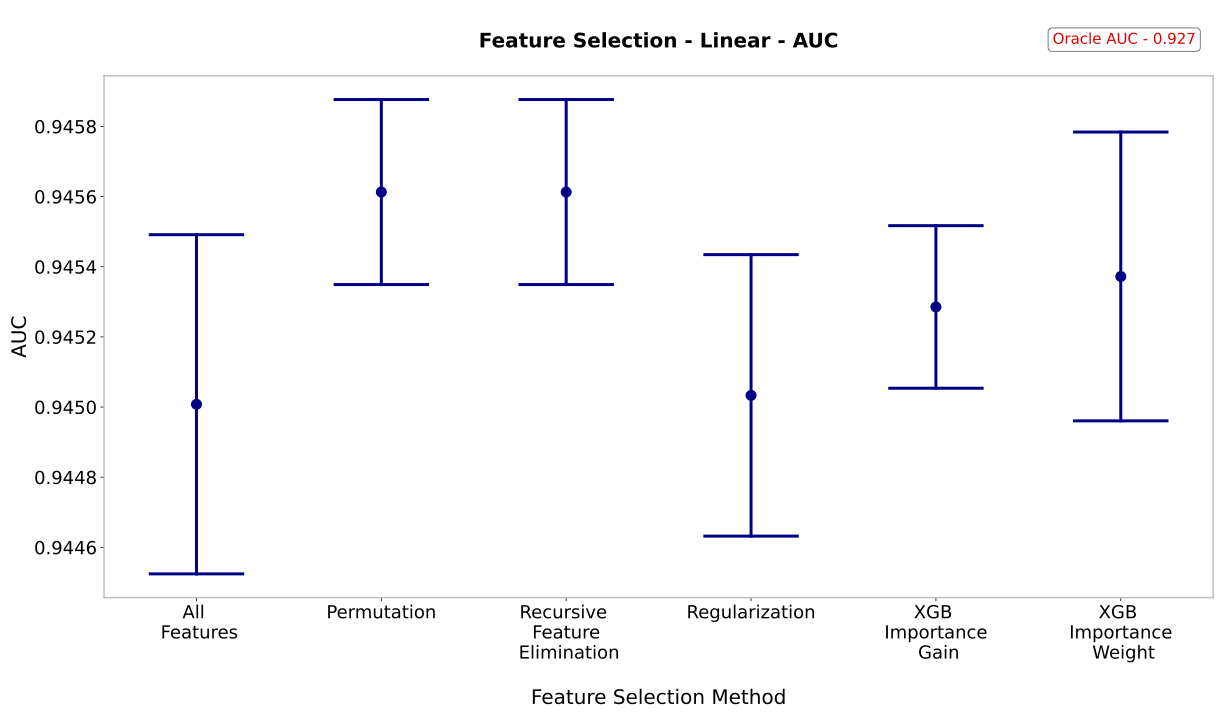}
        \caption{Comparison of feature selection methods for linear synthetic data set}
        \label{fig:featsel_linear2}
    \end{figure}
    We would like to note two exceptions to this otherwise general observation. The two correlation reduction methods using Pearson's and Spearman's rank correlation coefficients are exceptions for both the linear and jumpy GAM local data sets. We visualize the behavior for the linear data set in Figure \ref{fig:featsel_linear}. Note that Figure \ref{fig:featsel_linear} shows the same information that is seen in Figure \ref{fig:featsel_linear2}, but with the addition of the correlation coefficient reduction methods.
    \begin{figure}[h!]
        \centering
        \includegraphics[scale = 0.25]{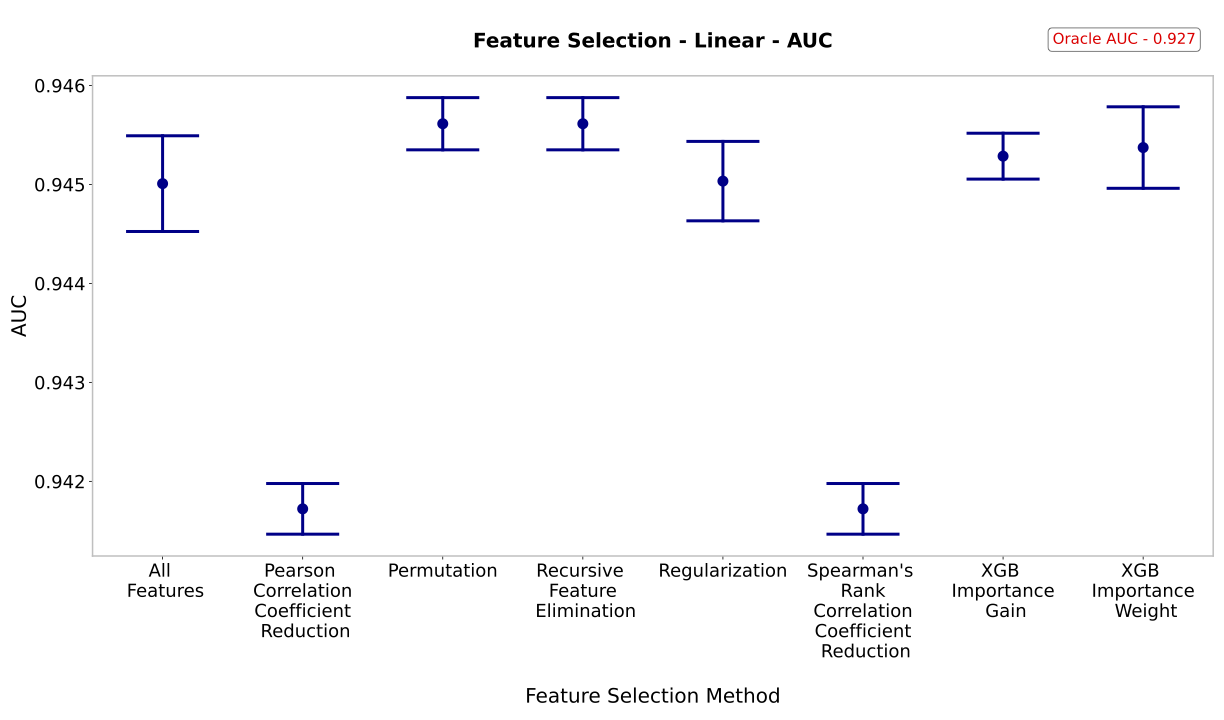}
        \caption{Comparison of feature selection methods for linear synthetic data set with addition of the correlation coefficient reduction methods}
        \label{fig:featsel_linear}
    \end{figure}
    We suspect that these show reduced performance due to the covariance matrix we enforced across the ten variables. We do not see similar behavior for GAM global, as there are fewer variables, all of which interact globally with one another to a certain extent. Since no variables interact with another variable in either the linear or jumpy GAM local data sets, any correlated pairs of variables are likely dropping one of the pair of variables, rather than keeping both. In fact, we see this behavior for the first pair of correlated features (i.e., features 0 and 1). Both correlation reduction methods drop feature 0 for jumpy GAM local data, but keep feature 1.
     \begin{figure}[h!]
        \centering
        \label{fig:featsel_corrvars}
        \includegraphics[scale = 0.39]{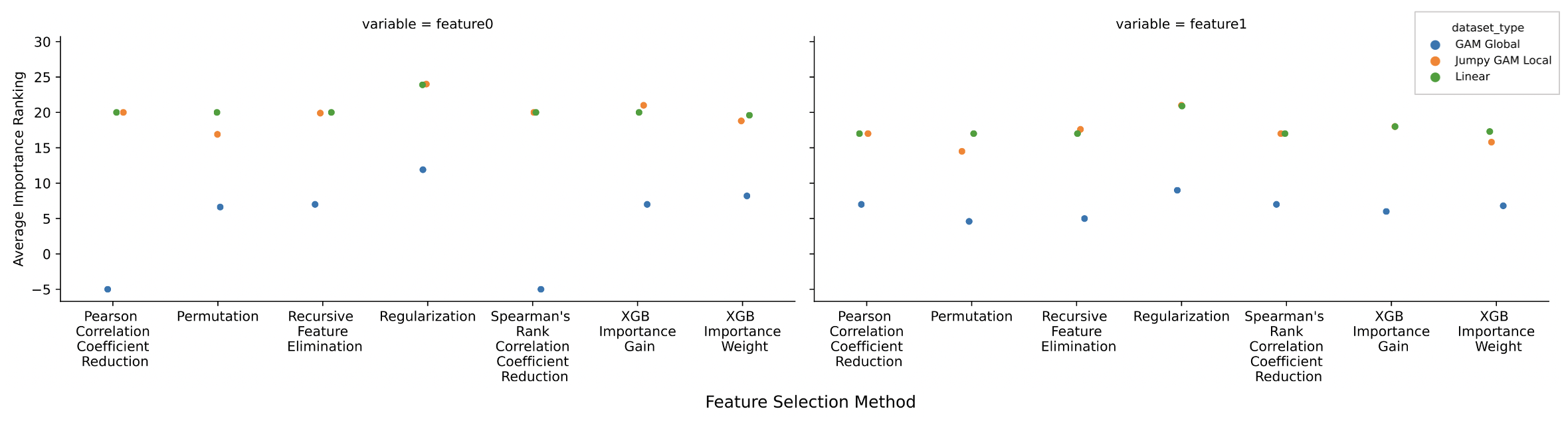}
        \caption{Comparison of average feature ranking for correlated variables 0 and 1 across feature selection methods for the three synthetic data sets}
    \end{figure}
    
    \item[Observation \ref{featselobs2}: Permutation-based feature importance has high variability relative to other methods for data that include local interactions, which suggests less stability among the identified feature subsets. We do not recommend using permutation-based feature importance as the preferred feature selection method. \label{obs:feat2}] \obs{featselobs2}
    We observe the highest performance variability for variable selection via permutation-based feature importance for data sets that contain local feature interactions (i.e., jumpy GAM local and the LC data sets). Figure \ref{fig:featsel_lc} shows increased performance variability for the LC data set.
    \begin{figure}[h!]
        \centering
        \includegraphics[scale = 0.25]{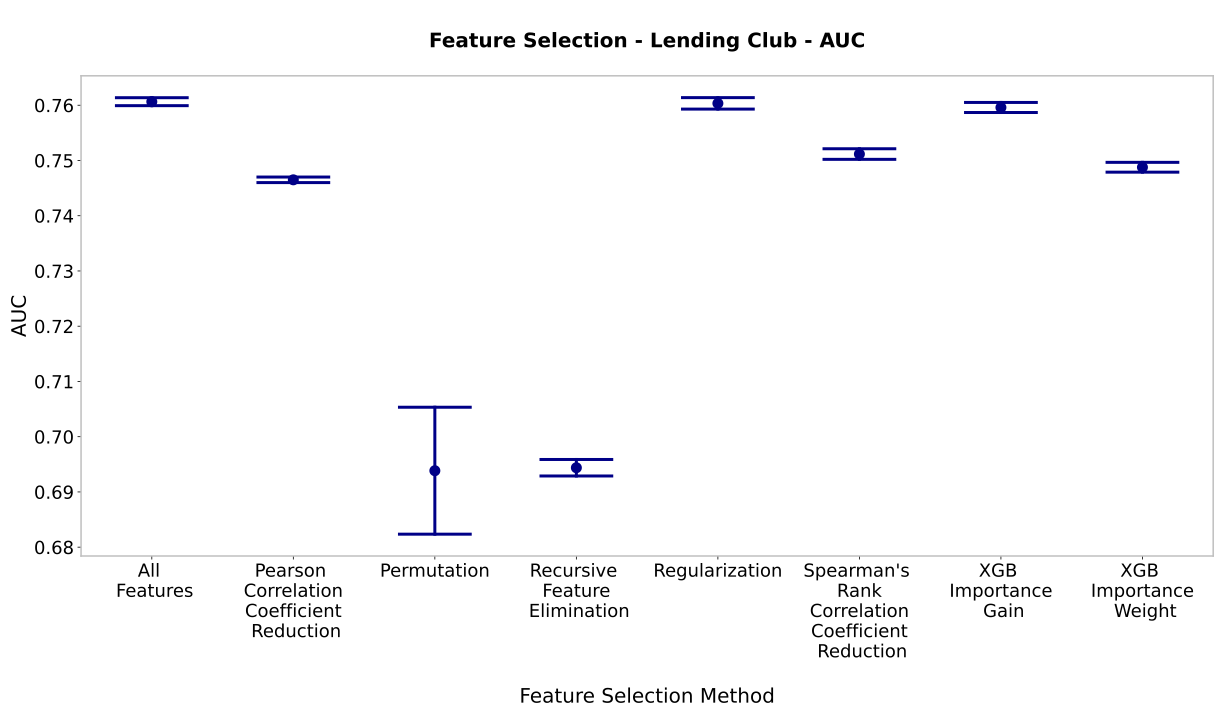}
        \caption{Comparison of feature selection methods for the Lending Club data set}
        \label{fig:featsel_lc}
    \end{figure}
    It is not observed in either linear or GAM global data sets, as the effect of permuted features are distributed globally in both cases. If the data set contains local interactions, the permutation will fundamentally change the underlying relationship and distribution of additive components within the data set, but is more dependent on the nature of the permutation. For example, if the permutation is more or less similarly distributed to the original data set, the perceived importance of the feature could appear to be less important than it actually is. On the other hand, if the permutation distribution is very different than the original data distribution, then the importance of the feature may be exaggerated. However, this will highly depend on both how the features are permuted as well as the order of the features being permuted, which could change the features identified as high importance.
    
    \quad A consequence of the permutation algorithm is the high potential to include variables of no importance or impact to the response variable. In these tests, permutation-based feature importance tended to include known noise variables (i.e., variables that were not included in the underlying functional form). Figure \ref{fig:featsel_noise13} shows permutation incorrectly giving noise variable 13 a high average importance ranking for jumpy GAM local.
    \begin{figure}[h!]
        \centering
        \includegraphics[scale = 0.35]{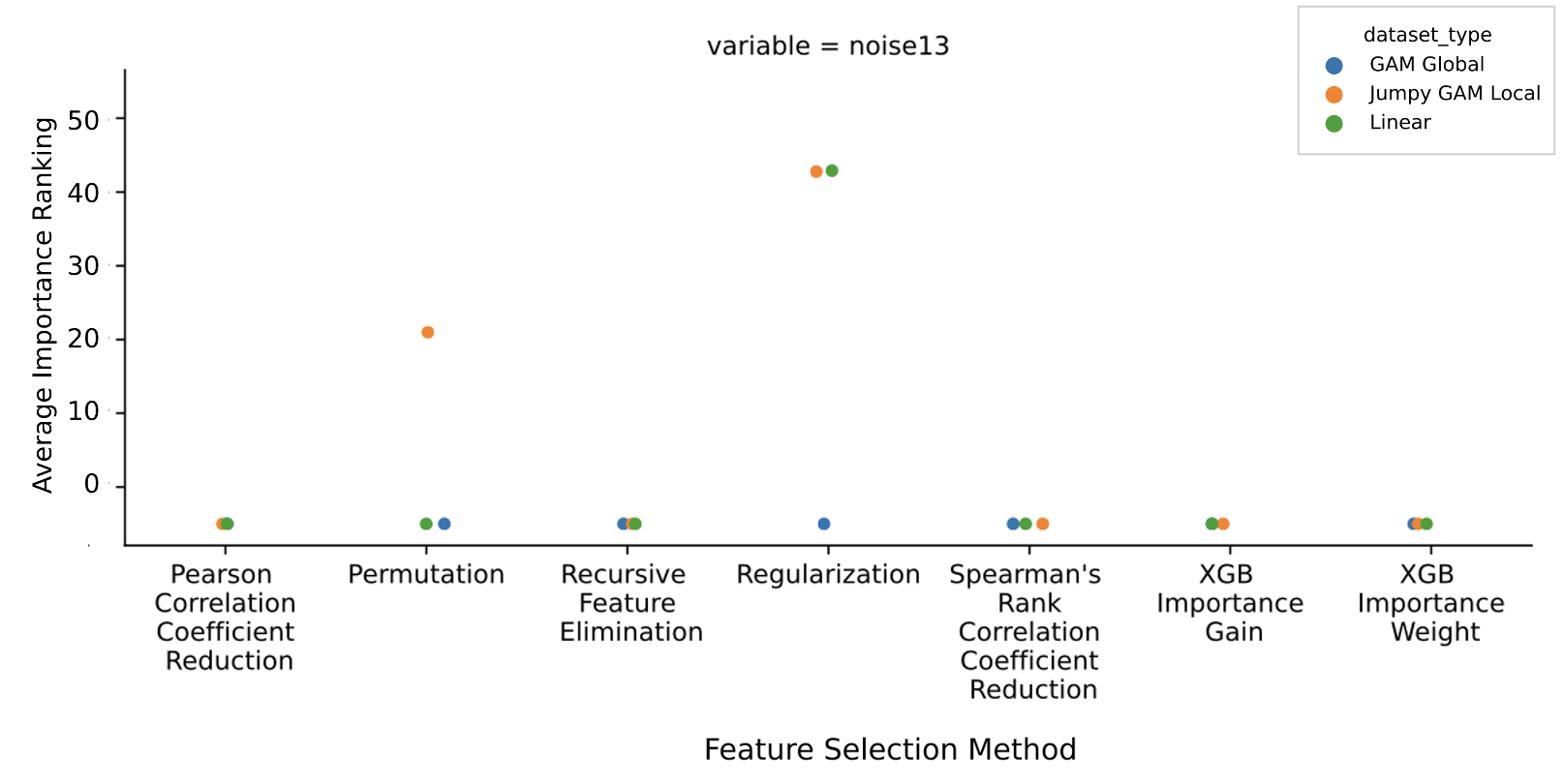}
        \caption{Comparison of average feature importance ranking for noise variable 13 for the three synthetic data sets}
        \label{fig:featsel_noise13}
    \end{figure}
    
    \item[Observation \ref{featselobs3}: Regularization often chose more variables than necessary, resulting in a more complex model with little to no gain in model performance. We do not recommend using regularization as the preferred feature selection method. \label{obs:feat3}] \obs{featselobs3}
    We artificially enforced all methods to select the same number of variables, with the exception of regularization. This is a key limitation of this study: artificially enforcing all methods to choose the same number of variables for consistent comparison. Due to how regularization operates, the regularization method was given free rein to include as many variables as it deemed appropriate. On average, we observed regularization choosing nearly as many variables as it possibly could (i.e., it chose all values included in the data set).
    
    \quad While regularization does not overtly show signs of overfitting, we do see hints of overfitting behavior. Here, we define overfitting as the gap between training and testing AUC. The larger the observed gap, the greater the evidence of overfitting. Figure \ref{fig:featsel_gap_jumpygamlocal} shows the larger than usual gap between the training and testing AUCs for regularization, which is on par with the performance seen when all features are included. In fact, they are quite nearly the same, as uncontrolled regularization almost always chose to use all available features, including noise (see Figure \ref{fig:featsel_noise13} for an example). This gap in training and testing AUCs is larger than any of the other observed gaps for the other feature selection methods.
    \begin{figure}[h!]
        \centering
        \includegraphics[scale = 0.25]{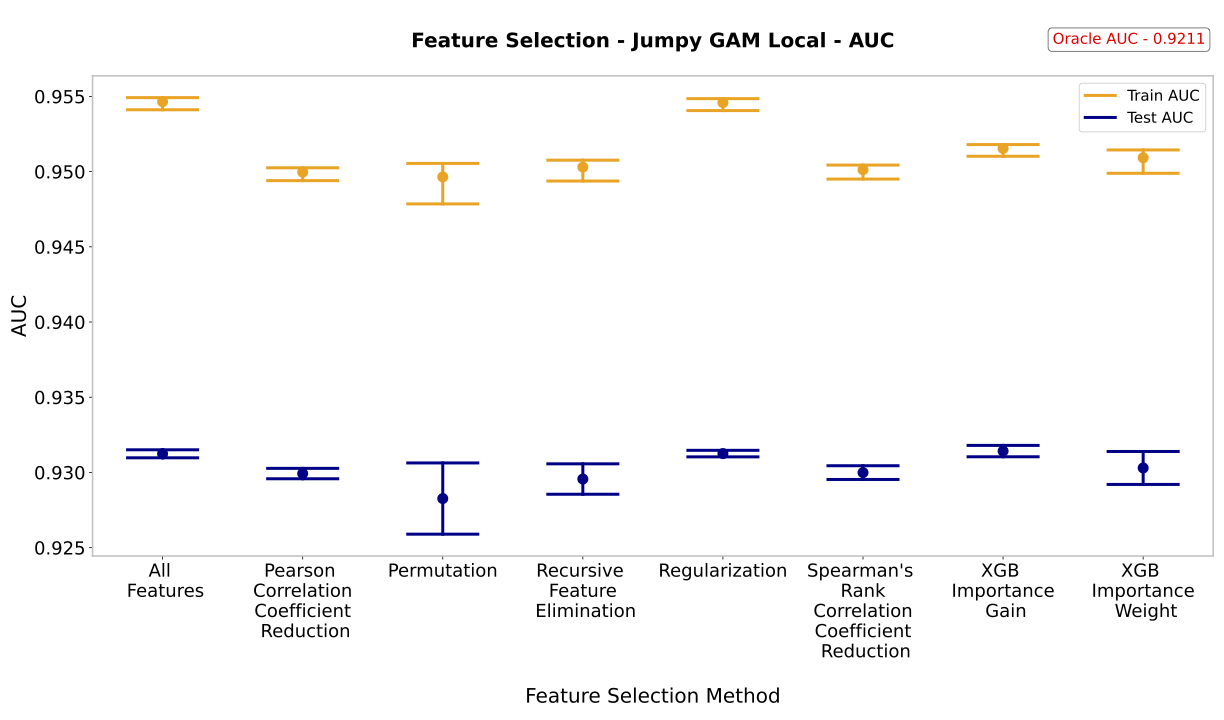}
        \caption{Comparison of the gap between training and testing AUC for feature selection methods for jumpy GAM local data set (correlation coefficient reduction methods were removed for better visualization)}
        \label{fig:featsel_gap_jumpygamlocal}
    \end{figure}
    \item[Observation \ref{featselobs4}: If relying on XGBoost's inherent feature importance, gain is preferred over weight. \label{obs:feat4}] \obs{featselobs4}
    Gain is the relative improvement in predicted response, whereas weight is the number of times a feature is used as splitting variable. We observed that features chosen using gain resulted in better model performance, both in number and variability (see Figure \ref{fig:featsel_lc} and Figure \ref{fig:featsel_jumpygamlocal2} for observed behavior in the LC and jumpy GAM local data sets, respectively). This behavior suggests that impact on response is a better indicator than frequency of feature usage.
    \begin{figure}[h!]
        \centering
        \includegraphics[scale = 0.25]{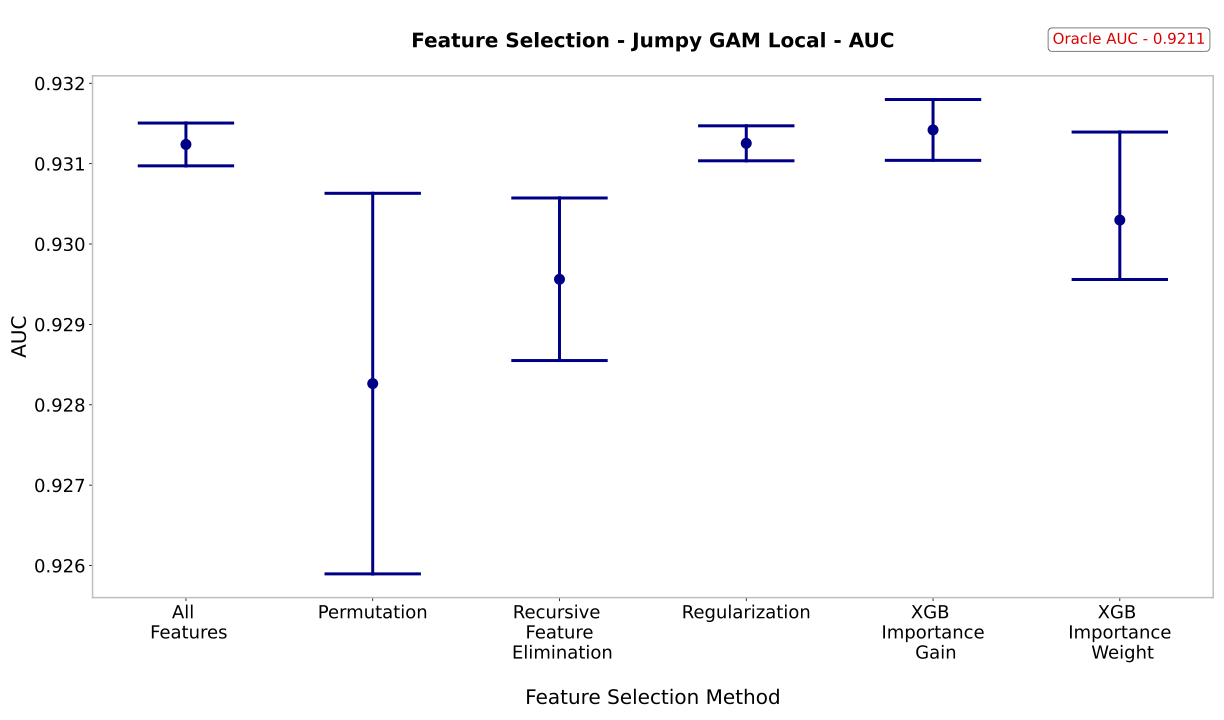}
        \caption{Comparison of feature selection methods for jumpy GAM local data set (correlation coefficient reduction methods were removed for better visualization)}
        \label{fig:featsel_jumpygamlocal2}
    \end{figure}
    \quad In fact, XGB importance by weight was the third most likely feature selection method to include noise variables. This indicates that noise variables were often used as splitting variables, without adding to model performance. They were likely filler. Figure \ref{fig:featsel_noise24} demonstrates this behavior for noise feature 24.
    \begin{figure}[h!]
        \centering
        \includegraphics[scale = 0.39]{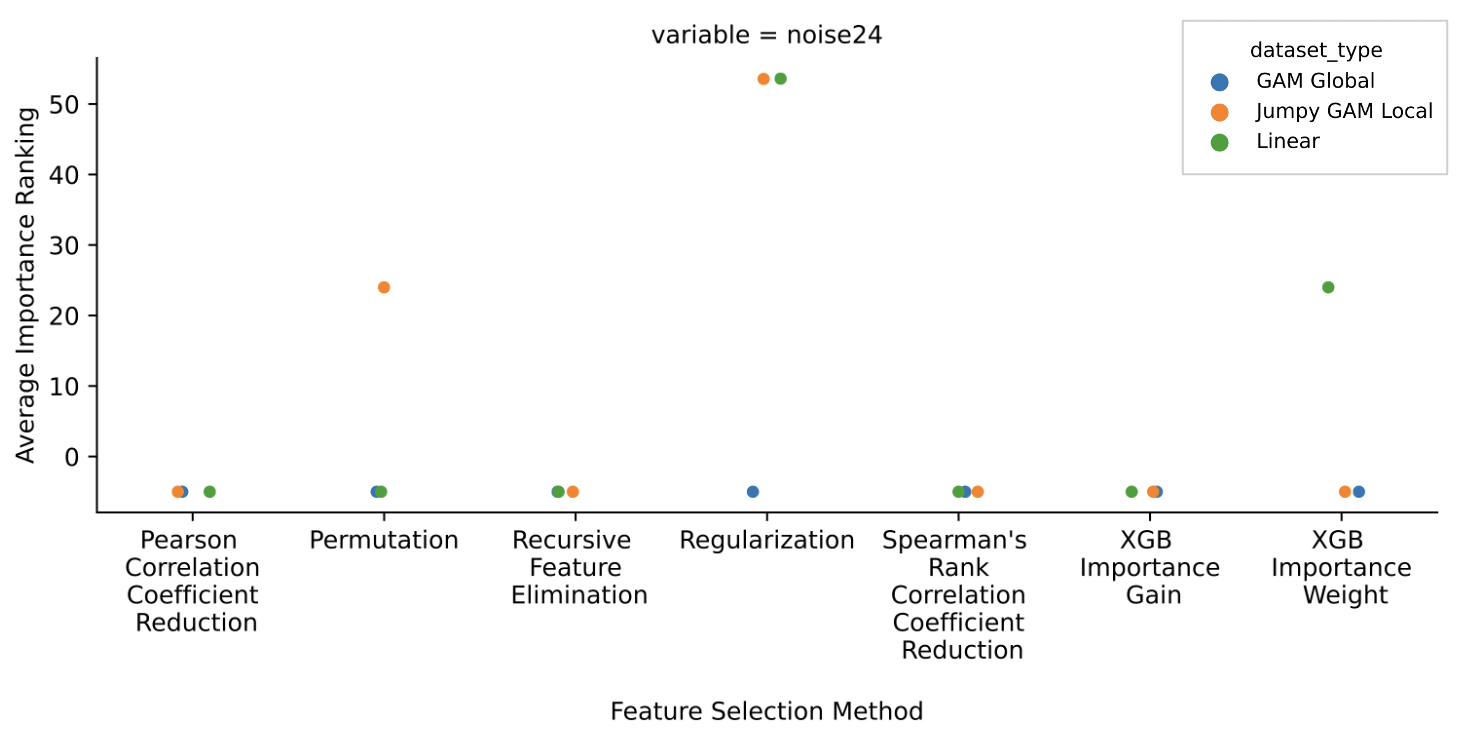}
        \caption{Comparison of average feature importance ranking for noise variable 24 for the three synthetic data sets}
        \label{fig:featsel_noise24}
    \end{figure}
    \item[Observation \ref{featselobs5}: XGB importance using gain shows the most consistent performance across a variety of data structures. \label{obs:feat5}] \obs{featselobs5}
    Across the four data sets considered in this study, XGB importance using gain shows consistently good performance relative to the other methods. Although it is not necessarily the ``best'' method in most cases, the performance variability shows dependable performance that ranks among the highest performers for all data sets. For example, Figures \ref{fig:featsel_linear2} and \ref{fig:featsel_gamglobal} exhibit this behavior. Additionally, in all data sets, this method shows no statistically significant difference in performance from the other highest ranking methodologies (see Figures \ref{fig:featsel_lc} and \ref{fig:featsel_jumpygamlocal2} for examples).
    \begin{figure}[h!]
        \centering
        \includegraphics[scale = 0.25]{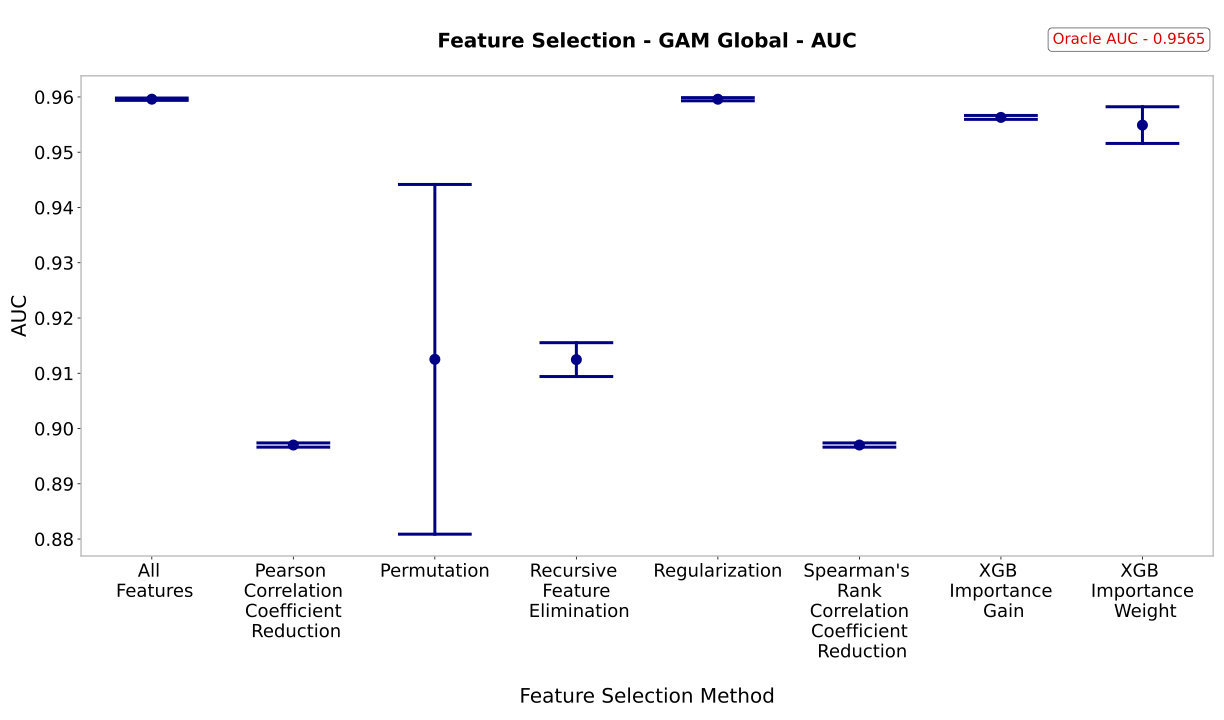}
        \caption{Comparison of feature selection methods for GAM global data set}
        \label{fig:featsel_gamglobal}
    \end{figure}

\end{description}

%% file: 03b_categorical.tex
\newcounter{catfindno}
\newcommand{\catfind}[1]{\refstepcounter{catfindno}\label{#1}}

\begin{description}[style = nextline]

    \item[Observation \ref{obscat1}: Frequency encoding performs poorly when data categories are structured and rigid, but works well in data with more complex categorical variable relationships. \label{obs:cat1}] \obs{obscat1}
    \quad Frequency encoding is the simplest method, as it adds the fewest additional variables to the data (i.e., a single additional variable). We found that this encoding approach works poorly with a very rigid, structured categorical data creation method. 
    
    \quad Our categories across the three synthetic data sets were forced to be mutually exclusive, necessitating the additional information provided by the three other encoding methods (i.e., OHE, binary encoding, and Helmert coding) that allows for much clearer distinction among the categories. While we observed this behavior in all three synthetic data sets, we use the results from the jumpy GAM local synthetic data set as an illustrative example (see Figure \ref{fig:catenc_jumpygamlocal}). With the use of binary, Helmert and one-hot encoding, near perfect AUC performance is achieved. Meanwhile, frequency encoding has a large drop in AUC performance comparatively.
    \begin{figure}[h!]
        \centering
        \includegraphics[scale = 0.25]{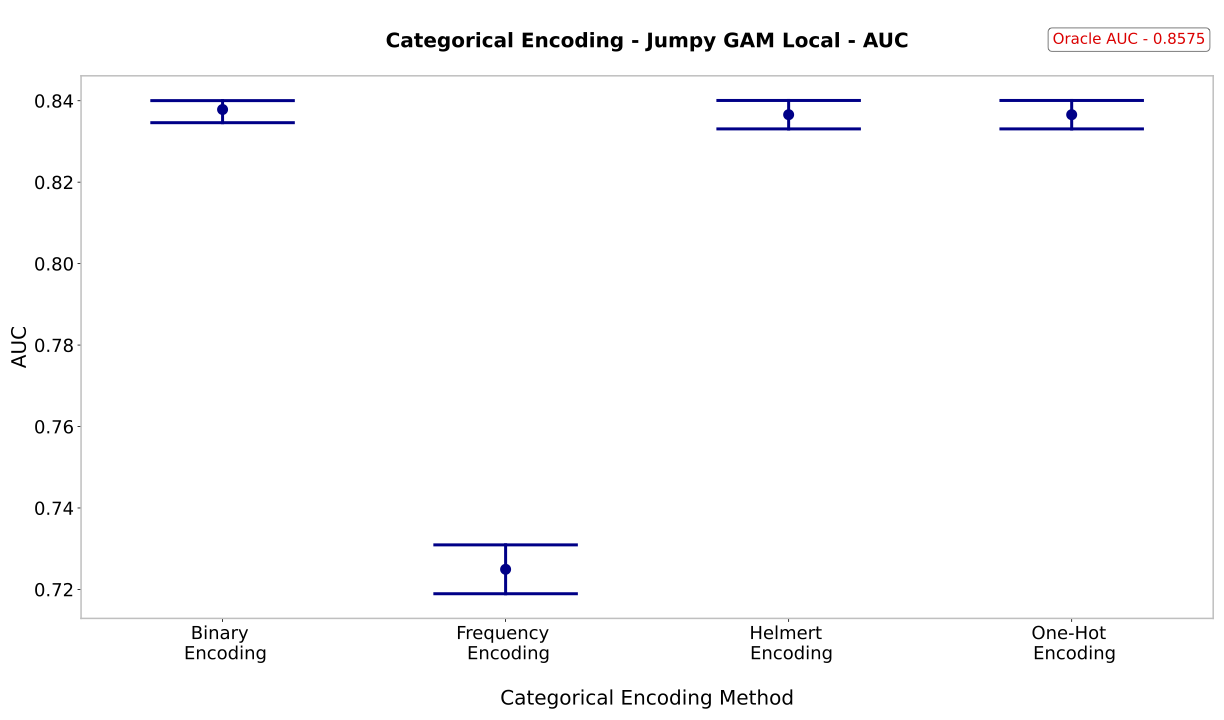}
        \caption{Comparison of categorical encoding methods for jumpy GAM local synthetic data set}
        \label{fig:catenc_jumpygamlocal}
    \end{figure}
    \quad We do not see the same phenomenon for the LC data, as there were ten categorical variables that were fighting to be recognized (see Figure \ref{fig:catenc_lc}). In fact, we see the complete opposite behavior, which can be explained by the addition of too many variables. Generally speaking, more variables does not necessarily result in better model performance, which was observed for the LC data. The contextual clues available across the categorical features from the frequency seemed to have allowed for improved model performance.
    \begin{figure}[h!]
        \centering
        \includegraphics[scale = 0.25]{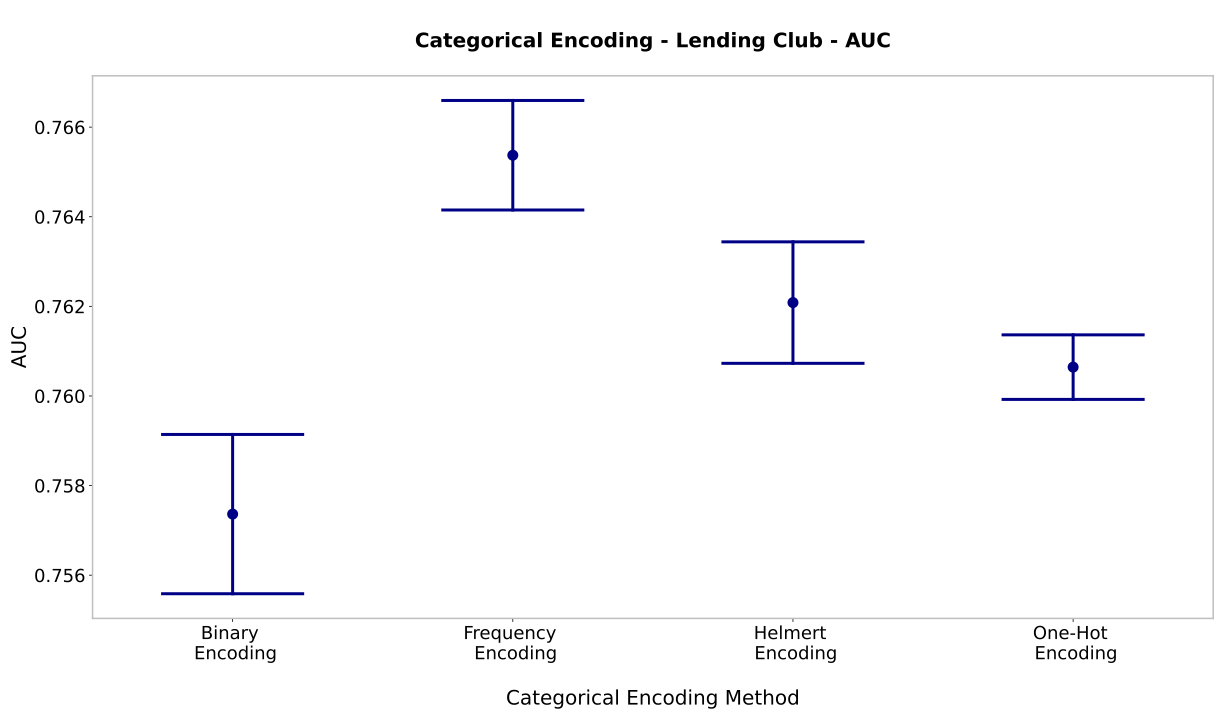}
        \caption{Comparison of categorical encoding methods for the Lending Club data set}
        \label{fig:catenc_lc}
    \end{figure}
    \item[Observation \ref{obscat2}: Helmert coding and OHE are related and hold categorical information in a similar way, such that the trained model results in identical predictions and encoded feature importance, for tree-based models. Either method are suitable options for categorical encoding, as both methods are comparable in performance. \label{obs:cat2}] \obs{obscat2}
    When looking at linear and jumpy GAM local data sets, we observed that the control group, which uses OHE, and the Helmert coding group had identical performance values. We investigated this further by pulling individual models from each group and comparing their feature importance weights. Table \ref{tab:ohe_helmert} displays the total number of times a variable was used for splitting within a tree (i.e., weight) made across all trees for a single model. Note this table focuses on the encoded categorical features, with the splits for continuous and noise features included in a limited capacity.
    
    \begin{table}[h!]
        \begin{center}
            {\setlength{\tabcolsep}{1.5em}
            \begin{tabular}{c c c} 
                 \toprule
                 \multirow{2}{*}{\textbf{Feature}} & \multicolumn{2}{c}{\textbf{Number of Splits}} \\
                 & OHE & Helmert Coding \\
                 \midrule
                 Feature 0 & 137  & 137 \\ 
                 Feature 1 & 147 & 147  \\ 
                 $\vdots$ & $\vdots$ & $\vdots$ \\ [0.6ex]
                 Categorical 1 & 128 & 262  \\ 
                 Categorical 2 & 134 & 127 \\ 
                 Categorical 3 & 127 & -- \\ 
                 \bottomrule
            \end{tabular}} \vspace{2mm}
        \end{center}
        \caption{Comparison of feature importance weights for models built using one-hot encoding and Helmert coding}
        \label{tab:ohe_helmert}
    \end{table}
    
    \quad OHE and Helmert coding have identical weights, with the exception of the categorical features created by their respective encoding method. OHE created three categorical features, with each feature denoting one of the three categories in the original categorical variable. Helmert coding created two categorical features, since each created feature inherently includes information to a reference category. In this case, the reference category information is plainly shown as the combination of Categorical 1 and Categorical 2 from OHE. Helmert coding is a way to simplify and reduce the number of added encoded features.

\end{description}

%% file: 03c_null_imputation.tex
\newcounter{nullfindno}
\newcommand{\nullfind}[1]{\refstepcounter{nullfindno}\label{#1}}

\begin{description}[style = nextline]

    \item[Observation \ref{nullobs1}: Missing indicator imputation shows the greatest performance across all data sets for both score value and variability, and is an excellent candidate for preferred null imputation method. \label{obs:null1}] \obs{nullobs1}
    Missing indicator imputation allows XGB to make the decision as to what values to impart for each missing value. The only information provided by this imputation method to the algorithm is whether the value is missing. Unlike the other imputation methods, missing indicator imputation does not give a predetermined, actual value. It allows the XGB to take advantage of its series of boosted trees to optimize what and how those missing observations are incorporated. While performance is not as good as if no data were missing, Figure \ref{fig:nullimp_linear} shows the marginally better performance of missing indicator imputation across all methods considered. Better performance is also observed for the LC data set in \ref{fig:nullimp_lc} and GAM global in Figure \ref{fig:nullimp_gamglobal}.
    \begin{figure}[h!]
        \centering
        \includegraphics[scale = 0.25]{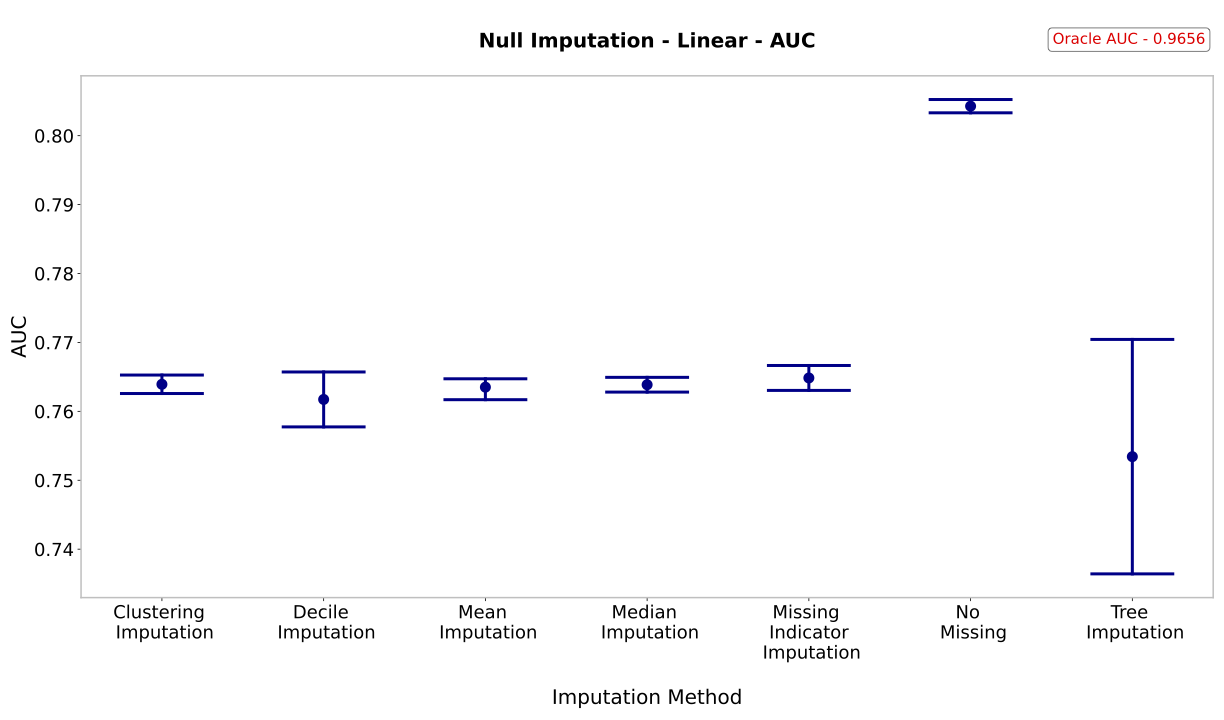}
        \caption{Comparison of null imputation methods for linear data set}
        \label{fig:nullimp_linear}
    \end{figure}
    
    \item[Observation \ref{nullobs2}: Single point imputation (i.e., decile, mean, and median imputation) all show very similar performance. \label{obs:null2}] \obs{nullobs2}
    \quad All single point imputation methods replace missing values with ``one'' value. Mean and median imputations are more similar among the three methods, as they are identical for symmetric distributions. Only in cases of skewed data will median imputation outperform mean imputation (e.g., Lending Club data in \ref{fig:nullimp_lc}). However, even in distributions with outliers or skew, median imputation only trivially outperforms mean imputation.
    \begin{figure}[h!]
        \centering
        \includegraphics[scale = 0.25]{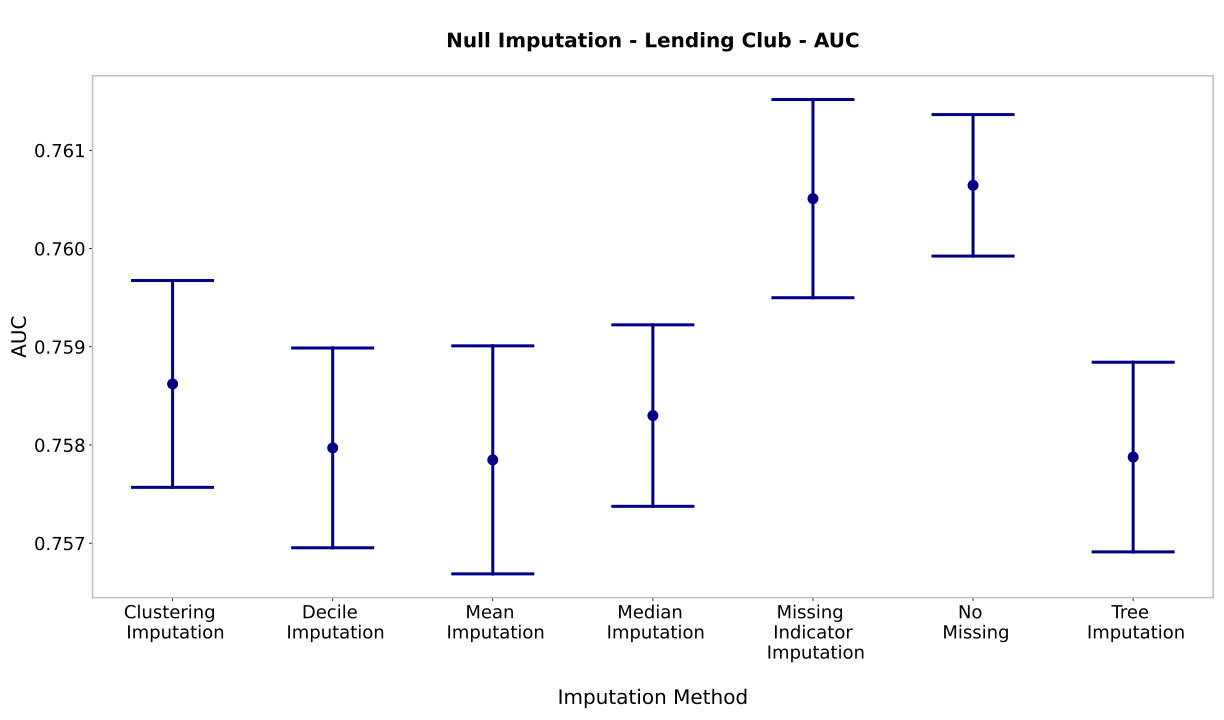}
        \caption{Comparison of null imputation methods for LC data set}
        \label{fig:nullimp_lc}
    \end{figure}
    \quad Decile imputation has the largest departure from the performances of mean and median imputation, as it is more granular, since it replaces the missing values with the mean within each decile. Decile imputation performs marginally better in some scenarios (e.g., GAM global in Figure \ref{fig:nullimp_gamglobal}), but not significantly. We can surmise in this case that the global interactions provide greater density of observations depending on the structure of the additive component. For example, $\beta_9\max(x_4, x_5)$ is determined by whether $x_4$ or $x_5$ is larger. Depending on how missing values were injected, there may be greater skew in their joint distribution. 
    \begin{figure}[h!]
        \centering
        \includegraphics[scale = 0.25]{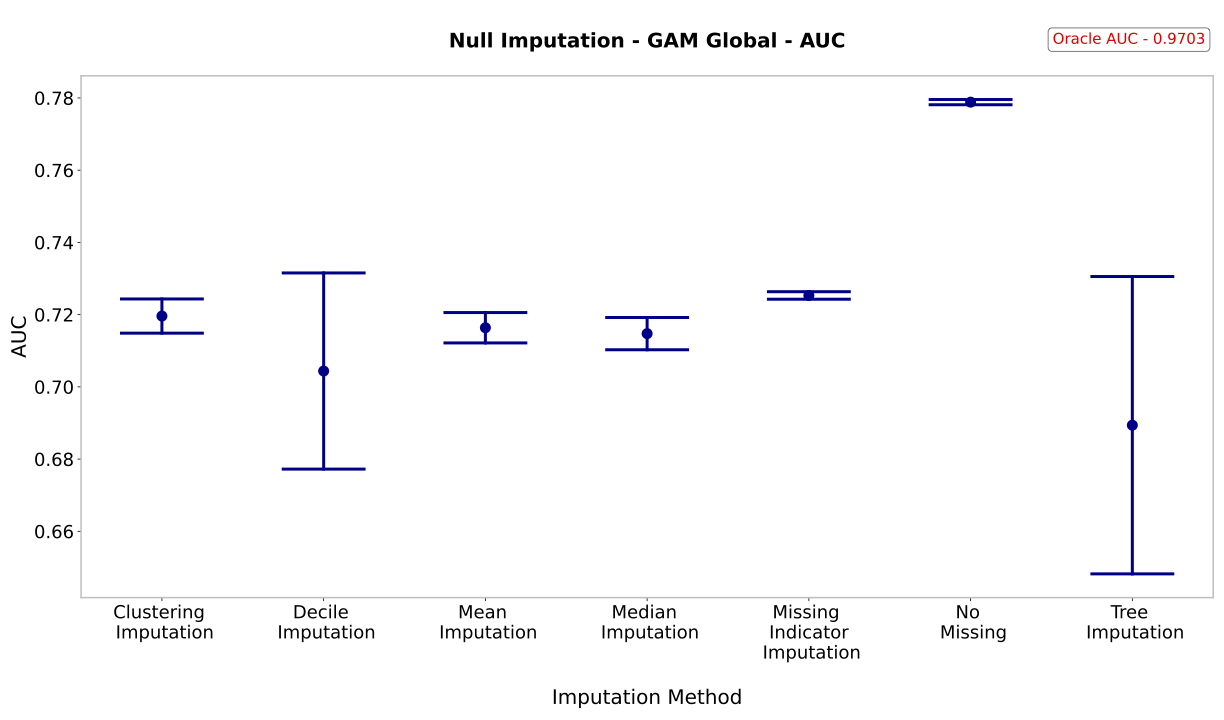}
        \caption{Comparison of null imputation methods for GAM global data set}
        \label{fig:nullimp_gamglobal}
    \end{figure}
    
    \item[Observation \ref{nullobs3}: Tree imputation has the most variable performance across data sets, and is not recommended for use. \label{obs:null3}] \obs{nullobs3}
    \quad The wide range of performance metrics seen for tree imputation suggests that it should not be the preferred imputation method. Our empirical results advise against the use of tree imputation in general, as other imputation methods provide as good or better performance for both score value and variability. This is supported by the algorithm of tree imputation, which builds a single decision tree for each feature to determine the imputed values. A single decision tree is widely recognized as being highly variable \citep{breiman2001random}.
    
    \quad The jumpy GAM local data set provides an excellent example of the highly variable nature, as see in Figure \ref{fig:nullimp_jumpygamlocal}. The data includes local interactions, which result in either a zero or non-zero value. This is quite striking in conjunction with the random choice of introducing missing values, as different missing rates among the local regions can highly influence imputed values and, as a result, model performance.
    \begin{figure}[h!]
        \centering
        \includegraphics[scale = 0.25]{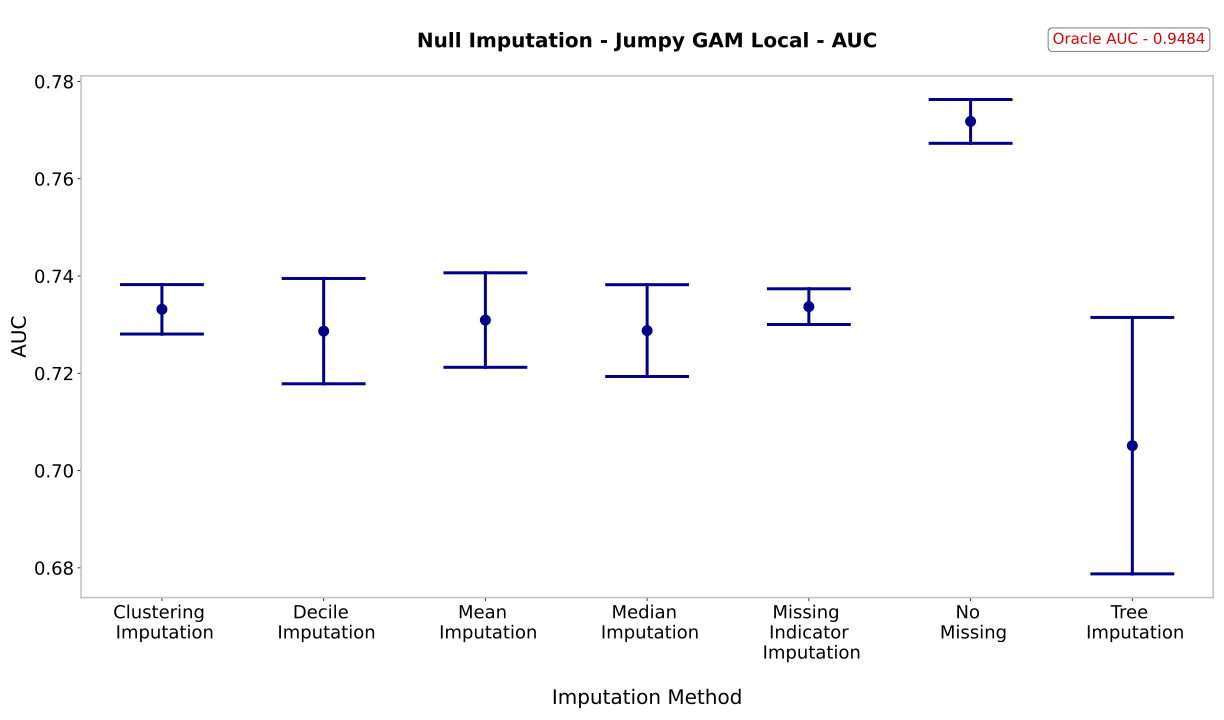}
        \caption{Comparison of null imputation methods for jumpy GAM local data set}
        \label{fig:nullimp_jumpygamlocal}
    \end{figure}

\end{description}

%% file: 04_limitations.tex
The scope of this study was limited. There are a number of avenues to explore in future work, we will list some of them in this section. 

\quad We assumed normal distributions for our continuous variables in our synthetic data, but there is opportunity to explore skewed or other distributions in future work. Additionally, we used a limited number of features in the synthetic data, which could be increased to provide comparisons in speed and computational complexity comparisons. The limitation in feature diversity extends to the types and number of categorical features included, as well.

\quad An additional limitation is the algorithm considered. We only examined these preprocessing methods with XGB models. We note that our findings and observations only pertain to XGB models. Other machine learning algorithms could be assessed as well (e.g., LightGBM, random forest, logistic regression, neural networks, etc.).

\quad This study only considered a binary classification model trained on tabular data, and future work may consider regression models, multiclass classification, and other data structures. 

\quad We would also like to note that our focus in this paper was on model performance, but the effect of these preprocessing methods on model interpretability and/or explainability is critical to understand. This is particularly crucial for fields such as financial services, where models are scrutinized by regulatory bodies to meet certain standards of model explainability. Future work could explore the impact of these preprocessing methods on \textit{post-hoc} interpretability methods, as well as models that are inherently interpretable, such as Explainable Boosting Machines \citep{nori2019interpretml}, interpretable deep neural networks \citep{sudjianto2021designing}, etc.

%% file: 05_conclusion.tex
Even though preprocessing is a major component of the modeling process, there are not many agreed upon standards or benchmarks to objectively guide modelers on how to optimize this step of model development. Most organizations rely on the experience of senior data scientists or engineers to choose the best techniques that fit the specific nuances and constraints of their data. In larger organizations, changes to recommended ``best-in-class'' methods may be difficult to percolate through the large population of quantitative analysts. Meanwhile, smaller organizations may not have the resources to make such determinations.

\quad To objectively deal with this roadblock, 2OS designed an experiment to look at a myriad of preprocessing techniques and benchmark their performance against each other. Our study has helped highlight preferred preprocessing methods, and identify less optimal ones. We approached this study in a pragmatic manner, with the intent of understanding the behavior of various preprocessing methods from an empirical standpoint. The goal was to provide a better and deeper understanding of the practical performance of popular feature selection, categorical encoding, and null imputation methods.

\quad We identified permutation-based feature importance and regularization as sub-optimal feature selection methods. In fact, permutation-based feature importance methods tend to have high variability when it comes to model performance. We would recommend avoiding these types of techniques. We also demonstrated that correlation coefficient reduction methods are not preferred approaches. At the same time, we observed and explained why XGB's feature importance via gain is a preferred method over weight. Overall, XGB's feature importance via gain showed the greatest consistency of performance among methods selected for experimentation. However, the choice of feature selection method is trivial for simple data structures, such as the synthetic linear data set.

\quad We also determined that frequency encoding can be the favored categorical encoding method, but only in certain circumstances. It is key to know and understand the complexity of the data structure before engaging in the categorical encoding step during preprocessing. In the case of frequency encoding, greater complexity in the data resulted in higher performance, in comparison to poorer performance for more rigidly structured data. We were also able to tangibly demonstrate the relationship between Helmert coding and OHE. Generally speaking, there was no universally preferred categorical encoding method.

\quad We established the dominance of missing indicator imputation among the imputation methods considered in this paper. We also illustrated the very similar behavior of single point imputation methods (i.e., decile, mean, and median imputations). Finally, we showed that tree imputation often resulted in highly variable and unstable model performance.

%% file: 06_appendix.tex
\section{Methodology}\label{sec:methodology}

\quad A test and control structure was used to run this experiment. The control group data sets had standard preprocessing techniques applied, whereas the test group data sets had experimental preprocessing techniques applied (with standard preprocessing done for areas unrelated to the test). Standard preprocessing techniques included enforcing column consistency between train and validation data sets, standardizing numeric columns, and one hot encoding categorical columns. The experimental methods used are described further in Appendix \ref{app:methods}. 

\quad After training and validation data sets for both control and test groups were appropriately preprocessed, an XGBoost classifier was trained and tuned. We used Bayesian hyperparameter optimization for tuning a few of the most important hyperparameters for our XGB classifier: gamma, learning rate, max depth, and number of estimators \citep{snoek2012practical}. Model performance was captured by calculating a number of metrics, but the final evaluation focused on using AUC. For our synthetic datasets, our benchmark was the ``oracle'' AUC. In this case, oracle refers to the best possible performance that the model could have if it knew the true, underlying probabilities. This is assessed based on the true response (i.e., probabilities) and the generated classes 0 and 1.

\quad In order to accurately gauge the impact each method had on model performance, control and test groups were run over multiple iterations to observe the degree of variance. In all cases, plots were created by calculating the mean and variance of the multiple iterations, and are shown as $\text{mean} \pm 2 \cdot \text{standard deviation}$. In some cases, the training and testing AUCs are plotted together to assess the severity of a model's overfit behavior.


\section{Data}\label{app:data}
Working with real-world data sets can be difficult when trying to benchmark performance for complex methods due to not knowing the true, underlying behavior of the data. Without controlling for the distribution of each feature and fully understanding the underlying data distribution, it is difficult to make generalized conclusions. In addition to using real-world data from Lending Club, we created three different types of synthetic data sets.

\subsection{Synthetic Data Generation}
By creating our own data sets, we are able to control the behavior within a single variable, as well as relationships across multiple variables. This means the impact and contribution of each variable is predetermined and can be anticipated. Each data set has a specific set of coefficients and each feature was generated based on a standard normal distribution. The response variable was generated from a predetermined data generating function. Each data set had a sample size of 250K.

\quad Each experiment had unique factors that made distinct characteristics of a synthetic data set more/less important. Multiple data generation procedures were implemented to allow each experiment to use a custom group of synthetic data sets. Taking into account the unique factors across experiments, the data generation procedure consisted of four major parts: 
\begin{enumerate}
    \item Assigning a functional form,
    \item Generating coefficients,
    \item Generating feature values, and
    \item Creating response variable.
\end{enumerate}

\quad There were three basic functional forms that were used to derive more complex forms across data sets from each experiment: linear, generalized additive model with global interactions (GAM global), and a jumpy generalized additive model with local interactions (jumpy GAM local), which were borrowed from \cite{liu2022performance}. Global interactions refer to functional forms that affect the entire feature space, while local interactions are only present in specific portions of the feature space.

\textbf{Linear additive model}: 
\begin{equation} 
    f(x) = \beta_1x_1 + ... + \beta_{10}x_{10}
    \label{eq:linear}
\end{equation}

\textbf{GAM with global interactions}: 
\begin{equation} 
    \begin{split}
        f(x) &= \beta_1|x_1| + \beta_2x_2^2 + \beta_3\log(|x_3| + 1) \\
        &+ \exp(\beta_4x_4) + (\beta_5|x_5| + 1)^{-1} + \beta_6x_1x_2 \\
        &+ \beta_7|x_1x_2x_3| + \beta_8\log(|x_3 + x_4 + x_5| + 1) \\
        &+ \beta_9\max(x_4, x_5) + \exp[\beta_{10}(x_5 - x_3)]
    \end{split}
    \label{eq:gamglobal}
\end{equation}

\textbf{Jumpy GAM with local interactions}: 
\begin{equation}
    \begin{split}
        f(x) &= \beta_1|x_1|\cdot I(|x_1| < 2) + \beta_2x_2^2\cdot I(x_2 > 1) \\
        &+ \beta_3\log(|x_3| + 1)\cdot I(|x_3| > 1) + \exp[\beta_4x_4]\cdot I(x_4 < 0) \\
        &+ (\beta_5|x_5| + 1)^{-1} + \beta_6\max(1, x_6 + x_7) + \beta_7I(x_7 < 1) \\
        &+ \beta_8I(|x_8| > 2) + \beta_9x_9\cdot I(x_9 < -1) + \beta_{10}\max(0,x_{10})
    \end{split}
    \label{eq:jumpygam}
\end{equation}

\quad \quad The function that defined each synthetic data set was made up of additive components that were weighted by a set of generated coefficients. The number of coefficients utilized depended on the structure and complexity of the specific type of function. Each coefficient $c_i$ was created by using NumPy’s random number generator, where $c_i \in [-3, 3]$ for $i = 1, ..., n$ \citep{harris2020array}.

\quad Additionally, the number of features created was dependent on the specific response function for each data set. A numeric feature can be represented by any set of scalar values with a fixed underlying distribution. The numeric features in our synthetic data sets follow a multivariate normal distribution. Correlation between feature pairs was also introduced by passing in a block-diagonal covariance matrix $\Sigma$ we created, where each pair had a correlation $r = 0.5$. For example, with 10 features, $\Sigma$ would be:
\begin{equation} \Sigma = 
    \begin{bmatrix}
    1 & r & 0 & 0 & 0 & 0 & 0 & 0 & 0 & 0 \\
    r & 1 & 0 & 0 & 0 & 0 & 0 & 0 & 0 & 0 \\
    0 & 0 & 1 & r & 0 & 0 & 0 & 0 & 0 & 0 \\
    0 & 0 & r & 1 & 0 & 0 & 0 & 0 & 0 & 0 \\
    0 & 0 & 0 & 0 & 1 & r & 0 & 0 & 0 & 0 \\
    0 & 0 & 0 & 0 & r & 1 & 0 & 0 & 0 & 0 \\
    0 & 0 & 0 & 0 & 0 & 0 & 1 & r & 0 & 0 \\
    0 & 0 & 0 & 0 & 0 & 0 & r & 1 & 0 & 0 \\
    0 & 0 & 0 & 0 & 0 & 0 & 0 & 0 & 1 & r \\
    0 & 0 & 0 & 0 & 0 & 0 & 0 & 0 & r & 1 \\
    \end{bmatrix}
    \label{eq:sigma}
\end{equation}

In cases of data sets with an odd number of features, the covariance matrix still exhibits a block-diagonal structure, but a variable independent of any other features.

\quad A categorical feature was also included in the null imputation and categorical encoding experiments. This was implemented by building each data set from the segment level, where each segment was defined by the categorical feature. Only a single categorical feature was used, so each distinct value represented a unique segment in the data. We expand on the details of our approach on implementing the categorical feature in the subsequent sections below.

\quad Additionally, ``noise'' features were introduced to the data set as way to deceive the XGB algorithm. These were numeric features that were produced using a random number generator, and did not contribute to the underlying function of the data set.

\quad The response variable for our data sets were calculated by mapping feature values to corresponding terms for a given function. The binary response $y$ was determined by using a Bernoulli distribution with probability $P(Y = 1 | x) = p(x)$, which is given by a sigmoid transformation (see Eq. \ref{eq:sigmoid}). To minimize bias in our models, each data set was inspected to ensure classes were balanced.

\begin{equation}
    f(x) = \log\left(\frac{p(x)}{1 - p(x)}\right)
    \label{eq:sigmoid}
\end{equation}

\quad As each group of preprocessing methods have different goals, the synthetic data sets used were customized to each preprocessing objective. We will go through the modifications for each group of methods below. Overall, 108 data sets were generated: 
\begin{itemize}
  \item Linear: 30 training sets, 3 validation (i.e., testing) sets, and 3 data sets for hyperparameter optimization
  \item GAM Global: 30 training sets, 3 validation sets, and 3 data sets for hyperparameter optimization
  \item Jumpy GAM Local: 30 training sets, 3 validation sets, and 3 data sets for hyperparameter optimization
\end{itemize}

\begin{description}[style = nextline]
\item[Null Imputation]
The null imputation experiment used the most basic procedure for generating synthetic data. There were three types of data sets, and their corresponding functional forms mimicked the basic forms described above in Equations \ref{eq:linear}, \ref{eq:gamglobal}, and \ref{eq:jumpygam}. 

\quad After each data set was created, three features were chosen to have missing values inserted in-place with a null-rate of 50\%. Tables \ref{tab:nullimp1} and \ref{tab:nullimp2} show toy examples of the original (no missing) and modified data sets (injected missing values). In this case, numeric features 2 and 3 were chosen as the designated variables to inject missing values. At a 50\% null-rate, this means that two of the four observations will become missing (i.e., NA).
\begin{table}[h!]
    \begin{center}
        \begin{tabular}{c c c c} 
             \toprule
             \textbf{Numeric Feature 1} & \textbf{Numeric Feature 2} & \textbf{Numeric Feature 3} & \textbf{Numeric Feature 4} \\ [0.5ex] 
             \midrule
             0.34 & 0.04 & 0.28 & 0.81  \\ 
             0.67 & 0.76 & 0.60 & 0.68 \\
             0.85 3 & 0.71 & 0.92 & 0.45 \\
             0.50 & 0.42 & 0.59 & 0.86 \\
             \bottomrule
        \end{tabular} \vspace{2mm}
    \end{center}
    \caption{Toy original data set with no missing values}
    \label{tab:nullimp1}
\end{table}

\begin{table}[h!]
    \begin{center}
        \begin{tabular}{c c c c} 
             \toprule
             \textbf{Numeric Feature 1} & \textbf{Numeric Feature 2} & \textbf{Numeric Feature 3} & \textbf{Numeric Feature 4} \\ [0.5ex] 
             \midrule
             NA & NA & 0.28 & 0.81 \\ 
             0.67 & NA & 0.60 & 0.68 \\
             NA 3 & 0.71 & NA & 0.45 \\
             0.50 & 0.42 & NA & 0.86 \\
             \bottomrule
        \end{tabular} \vspace{2mm}
    \end{center}
    \caption{Toy modified data set with injected missing values}
    \label{tab:nullimp2}
\end{table}

\quad Table \ref{tab:nullimp_summary} provides a summary of the structural components for the created data sets for the null imputation experiment. The categorical feature was chosen to include five categories or segments. Ten or five features contributed to the response, depending on the data set type, while five features were included noise variables. In total, either 15 or 10 features were included in the final data set.
\begin{table}[h!]
    \begin{center}
        \begin{tabular}{c c c c c} 
             \toprule
             \textbf{Data Set Type} & \textbf{Coefficients} & \textbf{Numeric Features} & \textbf{Noise Features} & \textbf{Segments} \\ [0.5ex] 
             \midrule
             Linear & 10 & 10 & 5 & 5 \\ 
             GAM Global & 10 & 5 & 5 & 5 \\
             Jumpy GAM Local & 10 & 10 & 5 & 5 \\
             \bottomrule
        \end{tabular} \vspace{2mm}
    \end{center}
    \caption{Summary description of each data set used in the null imputation experiment}
    \label{tab:nullimp_summary}
\end{table}

\item[Categorical Encoding]
The data sets produced in this experiment resembled the original functional forms. However, due to the importance of a categorical feature being introduced, they utilized indicator functions $I(\cdot)$ to act as ``gatekeepers'' in order to create larger additive sub-components in the data set. Since our models were XGB classifiers, this structure reinforced the necessity for our models to utilize the categorical feature to accurately predict the response variable. 

\quad Since the categorical encoding synthetic data were structurally different, we include the modified versions (see Equations \ref{eq:linear_ce}, \ref{eq:gamglobal_ce}, and \ref{eq:jumpygam_ce}) below. Note that these are still very similar to Equations \ref{eq:linear}, \ref{eq:gamglobal}, and \ref{eq:jumpygam}.

\textbf{Linear additive model (categorical encoding)}: 
\begin{equation} 
    \begin{split}
        f(x) &= I(x_{cat1} = 1) * (\beta_1x_1 + \beta_2x_2 + \beta_3x_3) \\
        &+ I(x_{cat2} = 1) * (\beta_4x_4 + \beta_5x_5 + \beta_6x_6) \\
        &+ I(x_{cat3} = 1) * (\beta_7x_7 + \beta_8x_8 + \beta_9x_9 + \beta_{10}x_{10})
    \end{split}
    \label{eq:linear_ce}
\end{equation}

\textbf{GAM with global interactions (categorical encoding)}: 
\begin{equation} 
    \begin{split}
        f(x) &= I(x_{cat1} = 1) * (\beta_1|x_1| + \beta_2x_2^2 + \beta_3\log(|x_3| + 1)) \\
        &+ I(x_{cat2} = 1) * (\exp(\beta_4x_4) + (\beta_5|x_5| + 1)^{-1} + \beta_6x_1x_2) \\
        &+ I(x_{cat3} = 1) * (\beta_7|x_1x_2x_3| + \beta_8\log(|x_3 + x_4 + x_5| + 1) \\
        &+  \beta_9\max(x_4, x_5) + \exp[\beta_{10}(x_5 - x_3)])
    \end{split}
    \label{eq:gamglobal_ce}
\end{equation}

\textbf{Jumpy GAM with local interactions (categorical encoding)}: 
\begin{equation}
    \begin{split}
        f(x) &= I(x_{cat1} = 1) * (\beta_1|x_1|\cdot I(|x_1| < 2) + \beta_2x_2^2\cdot I(x_2 > 1) + \beta_3\log(|x_3| + 1)\cdot I(|x_3| > 1)) \\
        &+ I(x_{cat2} = 1) * (\exp[\beta_4x_4]\cdot I(x_4 < 0) + (\beta_5|x_5| + 1)^{-1} + \beta_6\max(1, x_6 + x_7)) \\
        &+ I(x_{cat3} = 1) * (\beta_7I(x_7 < 1) + \beta_8I(|x_8| > 2) + \beta_9x_9\cdot I(x_9 < -1) + \beta_{10}\max(0,x_{10}))
    \end{split}
    \label{eq:jumpygam_ce}
\end{equation}

\quad Table \ref{tab:catenc_summary} provides a summary of the structural components for the created data sets for the categorical encoding experiment. The overall structure is very similar to the components described in Table \ref{tab:nullimp_summary}, with the exception of the number of segments or categories chosen. In this case, the categorical feature was chosen to include three segments. The remaining components are the same. Ten or five features contributed to the response, depending on the data set type, while five features were included noise variables. In total, either 15 or 10 features were included in the final data set.
\begin{table}[h!]
    \begin{center}
        \begin{tabular}{c c c c c} 
             \toprule
             \textbf{Data Set Type} & \textbf{Coefficients} & \textbf{Numeric Features} & \textbf{Noise Features} & \textbf{Segments} \\ \midrule
             Linear & 10 & 10 & 5 & 3 \\ 
             GAM Global & 10 & 5 & 5 & 3 \\
             Jumpy GAM Local & 10 & 10 & 5 & 3 \\
             \bottomrule
        \end{tabular} \vspace{2mm}
    \end{center}
    \caption{Summary description of each data set used in the categorical encoding experiment}
    \label{tab:catenc_summary}
\end{table}

\item[Feature Selection]
Similar to categorical encoding, the synthetic data sets that belonged the feature selection experiment expanded on the foundational functional forms in \ref{eq:linear_ce}, \ref{eq:gamglobal_ce}, and \ref{eq:jumpygam_ce}. The functions for these data sets were augmented to force our models to distinguish the relative importance among the features. This was accomplished by grouping terms in the functional form and using a coefficient to act as a weight of importance for that group. To avoid unnecessary complexity, the structure of terms across each group were identical, but the individual features values and coefficients were distinct. 

\quad Since the feature selection synthetic data were structurally different, we include the modified versions (see Equations \ref{eq:linear_fs}, \ref{eq:gamglobal_fs}, and \ref{eq:jumpygam_fs}) below. Note that these are still very similar to Equations \ref{eq:linear}, \ref{eq:gamglobal}, and \ref{eq:jumpygam}. The largest difference is the combination of three versions $g(x)$, $g'(x)$, and $g''(x)$ of the base functions that form the weighted sum (i.e., the response) $f(x)$.

\textbf{Linear additive model (feature selection)}: 
\begin{equation}
    \begin{split}
        g(x) &= \beta_1x_1 + ... + \beta_{10}x_{10} \\
        g'(x) &= \beta_{11}x_11 + ... + \beta_{20}x_{20} \\
        g''(x) &= \beta_{21}x_21 + ... + \beta_{30}x_{30} \\
        f(x) &= \beta_{31}\cdot g(x) + \beta_{32}\cdot g'(x) + \beta_{33}\cdot g''(x)
    \end{split}
    \label{eq:linear_fs}
\end{equation}

\textbf{GAM with global interactions (feature selection)}: 
\begin{equation} 
    \begin{split}
        g(x) &= \beta_1|x_1| + \beta_2x_2^2 + \beta_3\log(|x_3| + 1) \\
        &+ \exp(\beta_4x_4) + (\beta_5|x_5| + 1)^{-1} + \beta_6x_1x_2 \\
        &+ \beta_7|x_1x_2x_3| + \beta_8\log(|x_3 + x_4 + x_5| + 1) \\
        &+ \beta_9\max(x_4, x_5) + \exp[\beta_{10}(x_5 - x_3)] \\
        g'(x) &= \beta_{11}|x_6| + ... + \exp[\beta_{20}(x_{10} - x_8)] \\
        g''(x) &= \beta_{21}|x_{11}| + ... + \exp[\beta_{30}(x_{15} - x_{13})] \\
        f(x) &= \beta_{31}\cdot g(x) + \beta_{32}\cdot g'(x) + \beta_{33}\cdot g''(x)
    \end{split}
    \label{eq:gamglobal_fs}
\end{equation}

\textbf{Jumpy GAM with local interactions (feature selection)}: 
\begin{equation}
    \begin{split}
        g(x) &= \beta_1|x_1|\cdot I(|x_1| < 2) + \beta_2x_2^2\cdot I(x_2 > 1) \\
        &+ \beta_3\log(|x_3| + 1)\cdot I(|x_3| > 1) + \exp[\beta_4x_4]\cdot I(x_4 < 0) \\
        &+ (\beta_5|x_5| + 1)^{-1} + \beta_6\max(1, x_6 + x_7) + \beta_7I(x_7 < 1) \\
        &+ \beta_8I(|x_8| > 2) + \beta_9x_9\cdot I(x_9 < -1) + \beta_{10}\max(0, x_{10}) \\
        g'(x) &= \beta_{11}|x_{11}|\cdot I(|x_{11}| < 2) + .. + \beta_{20}\max(0, x_{20}) \\
        g''(x) &= \beta_{21}|x_{21}|\cdot I(|x_{21}| < 2) + .. + \beta_{30}\max(0, x_{30}) \\
        f(x) &= \beta_{31}\cdot g(x) + \beta_{32}\cdot g'(x) + \beta_{33}\cdot g''(x)
    \end{split}
    \label{eq:jumpygam_fs}
\end{equation}

Table \ref{tab:featsel_summary} provides a summary of the structural components for the created data sets for the feature selection experiment. The structural components in this case are very different than what seen for null imputation and categorical encoding. In this case, there was no categorical feature. Thirty or fifteen features contributed to the response, depending on the data set type, while 25 or 10 features were included noise variables. In total, either  or 10 features were included in the final data set.
\begin{table}[h!]
    \begin{center}
        \begin{tabular}{c c c c c} 
             \toprule
             \textbf{Data Set Type} & \textbf{Coefficients} & \textbf{Numeric Features} & \textbf{Noise Features} & \textbf{Segments} \\ [0.5ex] 
             \midrule
             Linear & 33 & 30 & 25 & 0 \\ 
             GAM Global & 33 & 15 & 10 & 0 \\
             Jumpy GAM Local & 33 & 30 & 25 & 0 \\
             \bottomrule
        \end{tabular} \vspace{2mm}
    \end{center}
    \caption{Summary description of each data set used in the feature selection experiment}
    \label{tab:featsel_summary}
\end{table}
\end{description}

\subsection{Lending Club Data}
Synthetic data allows for greater control for the data environment, but real-world data was still needed to properly explore the full effects of utilizing each method. Real-world data sets can be considered a more representative test of how a model will perform in production-level environments. Unlike synthetic data, the interactions or behaviors that features display can occur in unpredictable manners, with greater heterogeneity than would be seen in human-made data. This can expose a model to unforeseen circumstances or scenarios that would not have been seen otherwise. 

\quad For this experiment we used ``Lending Club Loan Data,'' which is a publicly available data set \citep{lendingclub2021}. Lending Club (LC) is a company that offers peer-to-peer personal loans. This data set contains anonymized loan data on customers that were issued loans between 2007 and 2015. It includes fields like credit score, debt-to-income ratio (DTI), term, months since last delinquency. Loan status is the target variable, which is categorized as ``Charged-off'' (i.e., account is considered as a loss) or ``Current.''

\quad For the LC data, in addition to the standard preprocessing that the experiment would handle, some initial preprocessing was performed to deal with unnecessary variables and potential sources of data leakage. This included variables with extremely high null rates (i.e., 99\%), identity variables (i.e., account number), and variables related to current delinquency (i.e., payment plan status). After these types of variables were removed, the data set was ready to be used in the experiment.

\quad Overall, the LC data set includes 2.26 million rows with 70 numeric and 10 categorical features. Of the 80 available features, 80\% of features have missing values.


\section{Methods}\label{app:methods}
We provide an overview of experimental methods utilized. The methods are broadly categorized into three groups:
\begin{enumerate}
    \item Feature selection,
    \item Categorical handling, and
    \item Null imputation
\end{enumerate}


\subsection{Feature Selection Methods}\label{app:feature}
Reducing the number of irrelevant input features is a key task for machine learning models to perform at optimal levels. By removing ``noisy'' features, the model will only utilize important variables to make predictions. Along with improvements in performance, using a subset of features can also help to reduce training time and improve model interpretability. The methods reviewed include:
\begin{itemize}
    \item Pearson correlation coefficient reduction,
    \item Spearman's rank correlation coefficient reduction,
    \item Variable selection based on XGBoost importance \citep{chen2016xgboost},
    \item Regularization via LASSO regression \citep{tibshirani1996regression},
    \item Variable selection based on permutation-based feature importance \citep{breiman2001random}, and
    \item Recursive feature elimination \citep{guyon2002gene}.
\end{itemize}

\begin{description}
\item[Pearson Correlation Reduction]
Pearson correlation reduction uses a two-step approach to remove unimportant features from the data. The first step deals with multicollinearity, while the second step looks at the correlation between a feature and the target variable. To deal with potential multicollinearity in the data, a correlation matrix is created that measures the observed correlation $r_{ab}$ of a feature $a$ with another feature $b$. This is done for every pair of features. 

\quad Any feature pairs that have a correlation above some specified threshold are candidates for removal. Only a single feature from the pair needs to be removed, so the feature in the pair that has the lowest correlation with the target variable $y$ dropped. This process is repeated for all feature pairs. After that, for the remaining list of features, the features with the highest correlation with the target variable (the top $N$ features) are kept. The formula for calculating the Pearson correlation coefficient is given in Eq. ref{eq:pearson}.

\begin{equation}
    r_{ab} = \frac{\sum_i\left(a_i - \bar{a}\right)\left(b_i - \bar{b}\right)}{\sqrt{\sum_i\left(a_i - \bar{a}\right)^2\sum_i\left(b_i - \bar{b}\right)^2}}
    \label{eq:pearson}
\end{equation}

\quad Explicitly dealing with multicollinearity is a nice benefit of this method, but having to calculate an entire covariance matrix measure correlation can become time consuming as the number of features increases. 

\item[Spearman's Rank Correlation Reduction]
Correlation reduction can also be performed using Spearman's rank correlation coefficients, rather than Pearson correlation coefficients. The formula for calculating Spearman's rank correlation coefficient is based on the Pearson correlation coefficient formula, but uses rank variables. Rather than using the original values, they are converted to ranks. This means that rank variables $R(a_i)$ and $R(b_i)$ are used in place of the original variables $a_i$ and $b_i$ in Eq. \ref{eq:pearson}. Similar to the Pearson correlation coefficient, dealing with multicollinearity can be computationally expensive as the number of features increases.

\item[Regularization – LASSO Regression]
LASSO (Least Absolute Shrinkage and Selection Operator) Regression (L1 regularization) is a linear model that adds a unique property to its cost function: a penalty term $\lambda$ that relates to the size of the coefficients in its equation \citep{tibshirani1996regression}. The higher the penalty term $\lambda$, the larger the constraint on the cost function. This results in an absolute reduction of the coefficient values. L1 regularization is used for feature selection because it will reduce the coefficients of less important features all the way to 0, which effectively removing them from the model. The cost function used for L1 regularization is given in Eq. \ref{eq:lasso}.
\begin{equation}
    f(x) = \sum_{i}\left(y_i- \sum_{j} x_{ij} \beta_j\right)^2 + \lambda\sum_{j}|\beta_j|
    \label{eq:lasso}
\end{equation}

\quad Unlike other feature selection methods that need to build multiple models to narrow down the top features, this method only builds one, which lead to relatively lower computing resources being leveraged. 

\item[XGBoost Importance]
XGBoost’s internal feature selection has the capability to output a list of importance scores for each score, from which the top $N$ important features can be identified \citep{chen2016xgboost}. XGBoost measures feature importance using three different mechanisms: gain, weight, and coverage. This experiment explored gain and weight. Gain is roughly calculated by looking at the increase in purity (i.e., ratio class in each node) of the children nodes when a feature is used as a split in the tree. If the observations in a parent node have balanced classes, and the resulting observations in the children nodes have very unbalanced classes, then that feature is considered to contribute to a larger gain. Weight is the number of splits that a feature had across all trees generated. The more times a feature is used to split across the trees, the larger the weight of that feature. 

\quad Five-fold cross validation, with 60\% train splits, was used to fit the model, and normalized feature importance scores were calculated. The scores were normalized by dividing the importance score for a single feature by the sum of scores for all features. The average score was then calculated across all folds, and the top $N$ features were returned.

\quad This method uses multiple metrics to measure how a feature contributes to the prediction made by the model. Depending on the type of features in the data set, this can be a positive or a negative. If the number of unique values for a feature is low, and the tree depth is high, then the number of times it can be used in a split in the tree will be limited and cause its weight metric to be dampened. This dampening can occur even if the feature is still important, which is certainly not ideal for the model developer.

\item[Permutation-Based Feature Importance]
Permutation-based feature importance measures importance by removing the predictive power of individual features and scoring the relative shift in performance of the model \citep{breiman2001random}. First, the entire data set is fitted to a model and scored. The performance of this model is stored as the benchmark for future calculations. Next, a single feature is chosen, and its values are randomly shuffled. New predictions are generated, and the new performance is measured. The new performance score should be lower than the benchmark, and the difference between the two is the feature importance score. The importance of the feature is directly related to the magnitude in difference between the two scores. This process is repeated for all features in the data set in isolation (only a single feature’s values are shuffled per step).

\quad The number of features scales with the number of models that need to be fitted for this method. Depending on the modelling technique used, computation can be expensive.

\item[RFE]
Recursive Feature Elimination (RFE) is a brute-force wrapper method that eliminates the lowest performing features from a data set in a step-wise manner \citep{guyon2002gene}. Being a wrapper method, RFE fits an ML model to a data set, scores the importance of each feature, and then removes the features with the lowest contribution. This is an iterative process that continues until the specified minimum number of features threshold is met. The size of the steps (i.e., the number of features removed per iteration) is another important parameter than can have a large effect on training time and performance. 

\quad RFE is a very simple algorithm that uses a large number of iterations to come up with a result. Because of this, computational cost can be expensive, especially as the number of features scale up.
\end{description}


\subsection{Categorical Handling Methods}\label{app:categorical}
ML algorithms only accept numerical inputs, so for an algorithm to learn the underlying patterns and relationships in the input data, every feature in that data set needs to be represented as a numeric variable. Due to this constraint, numerous encoding methods have been developed to translate categorical variables into numeric ones. 

\quad There are two different kinds of categorical variables: nominal and ordinal. Nominal variables have categories that share no intrinsic ordering between them (e.g., red, blue, yellow), while ordinal categories share a clear ordering between each category (e.g., small, medium, large). All the techniques we considered handle nominal variables,

\begin{itemize}
    \item One-hot encoding,
    \item Helmert coding \citep{sundstrom2010coding},
    \item Frequency encoding, and
    \item Binary encoding
\end{itemize}

A comparison of a select number of these methods was done by \cite{potdar2017comparative}.

\begin{description}
\item[One-Hot Encoding]
One-hot encoding (OHE) converts a categorical feature into $N - 1$ binary variables, where $N$ is the number of categories in the feature. Each new binary column corresponds to the original categorical feature where a ``1'' represents the presence of that category in the original feature. There are only $N - 1$ columns created since the final category is the ``base case” and is interpreted as the scenario when all other categories are 0. An example of OHE is given in Table \ref{tab:ohe}.

\begin{table}[h!]
    \begin{center}
        \begin{tabular}{c c c c} 
             \toprule
             \textbf{Category} & \makecell{\textbf{OHE} \\ \textbf{Category 1}} & \makecell{\textbf{OHE} \\ \textbf{Category 2}} & \makecell{\textbf{OHE} \\ \textbf{Category 3}} \\
             \midrule
             Category 1 & 1 & 0 & 0  \\ 
             Category 2 & 0 & 1 & 0 \\
             Category 3 & 0 & 0 & 1 \\
             \bottomrule
        \end{tabular} \vspace{2mm}
    \end{center}
    \caption{Example of one-hot encoding}
    \label{tab:ohe}
\end{table}

OHE is a straightforward technique for handling categorical variables, but it can have issues. When using this method for variables with high cardinality, OHE can become memory intensive and lead to high dimensionality within the data.

\item[Helmert Coding (Reverse)]
Helmert coding works by comparing a specific level of a categorical variable to the mean of the subsequent categories for that variable. A contrast is a linear combination (weighted sum) of statistics \citep{sundstrom2010coding}. Each contrast here is represented by the mean of the target variable for a specific level, subtracted by the mean of the means of the target variable for all categories that come after that level. For our implementation, we used a “reversed” form of Helmert coding in which previous categories are used as the comparison point instead of subsequent categories. See Table \ref{tab:helmertencoder} for an example.

\begin{table}[h!]
    \begin{center}
        \begin{tabular}{c c c c} 
             \toprule
             \textbf{Category} & \makecell{\textbf{Helmert} \\ \textbf{Level 1}} & \makecell{\textbf{Helmert} \\ \textbf{Level 2}} & \makecell{\textbf{Helmert} \\ \textbf{Level 3}} \\
             \midrule
             Category 1 & -0.5 & -0.33 & -0.25  \\ 
             Category 2 & 0.5 & -0.33 & -0.25 \\
             Category 3 & 0 & 0.66 & -0.25 \\
             Category 4 & 0 & 0 & 0.75 \\
             \bottomrule
        \end{tabular} \vspace{2mm}
    \end{center}
    \caption{Example of Helmert (Reverse) coding}
    \label{tab:helmertencoder}
\end{table}

\quad Ideally this method would be used for ordinal categorical variables since the resulting values are relative quantitative differences between categories, but it can still be utilized for nominal variables. The only discrepancy is that each value will be evaluated as a magnitude, and not discretely.

\item[Frequency Encoding]
Frequency encoding is a technique that replaces the literal value of a category with the probability of that category occurring within the data set. For example, if a data set with 100 rows had a categorical variable with 3 unique categories, where category 1 has 25 occurrences, category 2 has 60 occurrences, and category 3 has 15 occurrences, the newly created adjacent ``frequency'' column would use 0.25, 0.60, and 0.15 as the new respective corresponding values. For a categorical feature with 2 unique values, where category 1 has 72 occurrences and category 2 has 28 unique occurrences, the newly created adjacent frequency column would use 0.72 and 0.28 as the new respective corresponding values.See Table \ref{tab:freqencoding} for an example.

\begin{table}[h!]
    \begin{center}
        \begin{tabular}{c c c} 
             \toprule
             \textbf{Feature} & \textbf{Category} & \textbf{Frequency} \\ \midrule
             \multirow{3}{*}{Categorical Feature 1} & Category 1 & 0.25 \\ [0.5ex] 
             & Category 2 & 0.60 \\ [0.5ex] 
             & Category 3 & 0.15 \\ [0.5ex] \midrule
             \multirow{2}{*}{Categorical Feature 2} & Category 1 & 0.72 \\ [0.5ex] 
             & Category 2 & 0.28 \\ [0.5ex] \bottomrule
        \end{tabular} \vspace{2mm}
    \end{center}
    \caption{Example of frequency encoding}
    \label{tab:freqencoding}
\end{table}

\quad A big advantage to this technique is that it is simple and cost efficient to implement, while also keeping the size of the feature space constant. One major drawback from this method is the scenario when duplicate probabilities occur because the resulting values no longer serve as a way to differentiate unique segments.

\item[Binary Encoding]
Binary encoding represents each unique category as binary code across columns in the data set. In application, the categorical variable first needs to be transformed into an ordinal variable (no intrinsic relationship, simply need numeric values). The ordinal value for that category is then translated into binary code by utilizing the least number of columns necessary to represent every category in the feature. See Table \ref{tab:binary} for an example.

\begin{table}[h!]
    \begin{center}
        \begin{tabular}{c c c} 
         \toprule
         \textbf{Categorical Feature 1} & \textbf{Binary Categorical 1} & \textbf{Binary Categorical 2} \\ 
         \midrule
         Category 1 & 0 & 1 \\ 
         Category 2 & 1 & 0 \\
         Category 3 & 1 & 1 \\ \bottomrule
        \end{tabular} \vspace{2mm}
    \end{center}
    \caption{Example of binary encoding}
    \label{tab:binary}
\end{table}

\quad Binary encoding is a memory efficient method for dealing with categorical variables that have high cardinality. It is able to represent many categories with just a few columns being added to the data set. 
\end{description}


\subsection{Null Imputation Methods}\label{app:null}
Null or missing values can occur when there is absent information within a data set. Commonly referred to with ``NA'', null values are a common obstacle in machine learning for a few reasons. First, most machine learning algorithms are not able to handle missing values appropriately and will fail to fit a model if null values are present. Second, the presence of missing values may have negative impacts on model performance if those observations contain valuable information that is not present in the rest of the data set. Ignoring or deleting observations that have missing values may not be a luxury the analysis can afford (i.e., limited data) and can result in biasing the data set due to eliminating underlying behavior. There is debate on which methods are best for optimizing model performance. We will discuss several techniques in this section:

\begin{itemize}
    \item Mean imputation,
    \item Median imputation,
    \item Missing indicator imputation,
    \item Decile imputation,
    \item Clustering imputation \citep{lloyd1982least}, and
    \item Decision tree imputation
\end{itemize}

To demonstrate the above methods, we will use a toy data set, shown in Table \ref{tab:impdata}.

\begin{table}
    \begin{center}
        \begin{tabular}{c || c} 
             \toprule
             \textbf{Numeric Feature 1} & \textbf{Numeric Feature 2} \\ 
             \midrule
             100 & NA \\ 
             200 & 0.30 \\ 
             150 & 0.60 \\ 
             NA & 0.25 \\ 
             300 & 0.80 \\ 
             NA & 0.65 \\ \bottomrule
            \end{tabular} \vspace{2mm}
        \caption{}
        \label{tab:impdata}
    \end{center}
\end{table}

\begin{description}
\item[Mean Imputation]
This is a simple imputation method that replaces missing values for a specific feature with the mean of all non-missing values in the same feature. The toy data set in Table \ref{tab:impdata} shows missing values that are imputed via mean imputation, resulting in the data given in Table \ref{tab:meanimp}.

\begin{table}[h!]
    \begin{center}
        \begin{tabular}{c || c} 
             \toprule
             \textbf{Numeric Feature 1} & \textbf{Numeric Feature 2} \\ \midrule
             100 & 0.52 \\  
             200 & 0.30 \\ 
             150 & 0.60 \\ 
             187.5 & 0.25 \\ 
             300 & 0.80 \\ 
             187.5 & 0.65 \\ \bottomrule
        \end{tabular} \vspace{2mm}
    \end{center}
    \caption{Example of mean imputation of the toy data set from Table \ref{tab:impdata}}
    \label{tab:meanimp}
\end{table}

\quad While mean imputation is simple and easy to implement, the method does have some drawbacks. First, if the data is not normally distributed, using the mean to impute missing values may cause a change in the underlying distribution of the data. Additionally, if the percent of data missing is large enough, then mean imputation may lead to an underestimation of the feature’s variance. 

\item[Median Imputation]
Similar to mean imputation, this method replaces all missing values in a feature with the median value of all non-missing values in the same feature. The toy data set in Table \ref{tab:impdata} shows missing values that are imputed via median imputation, resulting in the data given in Table \ref{tab:medimp}.

\begin{table}[h!]
    \begin{center}
        \begin{tabular}{c || c} 
             \toprule
             \textbf{Numeric Feature 1} & \textbf{Numeric Feature 2} \\ \midrule
             100 & 0.6 \\
             200 & 0.30 \\
             150 & 0.60 \\ 
             175 & 0.25 \\ 
             300 & 0.80 \\ 
             175 & 0.65 \\ \bottomrule
        \end{tabular} \vspace{2mm}
        \caption{Example of median imputation of the toy data set from Table \ref{tab:impdata}}
        \label{tab:medimp}
    \end{center}
\end{table}

\quad Median imputation is a simple and fast approach for dealing with null values. Similar to mean imputation, if the percent of missing data is high enough, there may be a reduction in the feature’s variance. 

\item[Missing Indicator Imputation]
This is a simple technique where a binary feature is created to indicate whether the corresponding feature has a missing value present. The toy data set in Table \ref{tab:impdata} shows missing values that are imputed via missing indicator imputation, resulting in the data given in Table \ref{tab:missimp}.

\begin{table}[h!]
    \begin{center}
        \begin{tabular}{c c || c c} 
             \toprule
             \multicolumn{2}{c ||}{\textbf{Numeric Feature 1}} & \multicolumn{2}{c}{\textbf{Numeric Feature 2}} \\
             \textit{Original} & \textit{Missing Indicator} & \textit{Original} & \textit{Missing Indicator} \\ \midrule
             100 & 0 & -9999 & 1 \\ 
             200 & 0 & 0.30 & 0 \\ 
             150 & 0 & 0.60 & 0 \\ 
             -9999 & 1 & 0.25 & 0 \\ 
             300 & 0 & 0.80 & 0 \\ 
             -9999 & 1 & 0.65 & 0 \\ \bottomrule
        \end{tabular} \vspace{2mm}
        \caption{Example of missing indicator imputation of the toy data set from Table \ref{tab:impdata}}
        \label{tab:missimp}
    \end{center}
\end{table}

\quad The advantage of the indicator column is that it is able to highlight differentiating behavior that the presence of a missing value represents, but that is only if the missing value is not simply a random occurrence. Another drawback can be the scenario where many features have small missing rates, which increases dimensionality.

\item[Decile Imputation]
This is an imputation method that takes advantage of the relationship between the target variable and any missing features. For classification, the goal is to group the observations in the feature into percentile groups and create an additional group for missing values. We focus on deciles. For each group of observations (including the missing group), the probability of the target class occurring is calculated. The group with a target probability closest to the target probability of the missing group is chosen, and the median value for that group is used as the imputation value for that feature.

\quad This is a flexible method that can be used for either numerical or categorical variables depending on the statistic (e.g., mean, median, mode, etc.) used for the imputation value. The biggest advantage of this approach is that it utilizes the relationship of the target variable to try and associate missing values with a corresponding segment in the feature. One drawback of this method is that it assumes missing values have a strong relationship with other segments in the data, but if that is not the case (i.e., missing values are due to data entry errors), the feature becomes biased towards the segment chosen.

\item[Clustering Imputation]
Clustering imputation assigns clusters to every observation in the data set. Each feature with a missing value is then isolated and the average value of a cluster of observations within a feature is mapped. Clusters are identified using $k$-means clustering \citep{lloyd1982least}. The mean value for a cluster is then used as the imputation value for any missing values assigned to the same cluster in that feature. The toy data set from Table \ref{tab:clustimpdat} shows missing values, including data clusters, that are imputed via clustering imputation, resulting in the data given in Table \ref{tab:clustimp}.

\begin{table}[h!]
    \begin{center}
        \begin{tabular}{c c c} 
             \toprule
             \textbf{Numeric Feature 1} & \textbf{Numeric Feature 2} & \textbf{Cluster} \\  \midrule
             100 & NA & 1 \\ 
             200 & 0.30 & 2 \\
             150 & 0.60 & 2 \\
             NA & 0.25 & 2 \\
             300 & 0.80 & 1 \\
             NA & 0.65 & 1 \\
             \bottomrule
        \end{tabular} \vspace{2mm}
    \caption{Toy data set to demonstrate clustering imputation}
    \label{tab:clustimpdat}
    \end{center}
\end{table}

\begin{table}[h!]
    \begin{center}
        \begin{tabular}{c c c} 
            \toprule
             \textbf{Numeric Feature 1} & \textbf{Numeric Feature 2} & \textbf{Cluster} \\ \midrule
             100 & 0.725 & 1 \\ 
             200 & 0.30 & 2 \\
             150 & 0.60 & 2 \\
             175 & 0.25 & 2 \\
             300 & 0.80 & 1 \\
             200 & 0.65 & 1 \\
            \bottomrule
        \end{tabular} \vspace{2mm}
    \caption{Example of clustering imputation of the toy data set from Table \ref{tab:clustimpdat}}
    \label{tab:clustimp}
    \end{center}
\end{table}
\quad The advantage of clustering imputation is it utilizes information outside of the feature the missing value appears in. It assigns a missing value to the appropriate segment by using information from the rest of the data set, and then uses internal information about the feature to assign a specific value. Since this is a technique that uses a model to extract information from the data, it is high on the spectrum of computational cost relative to other methods.  

\item[Decision Tree Imputation]
This is another technique that uses a model to extract more information out of the data to determine an imputation value. Like decile imputation, the goal is to create unique segments or groups for an individual group and compare the average target value to a grouping of the missing value’s average target. 

\quad This technique uses a decision tree with a single feature to create the groupings. The decision tree is a classification and regression tree or CART \citep{breiman1984cart}. A $t$-test is performed to determine the similarity between each leaf node and the missing group’s average target values. The median feature value of the leaf node with the smallest $t$-score is used as the imputation value for that feature, and the process is repeated across all features. 

\quad Note that using a decision tree adds the overhead of training a model for every feature.
\end{description}

%% file: 01_main.bbl
\begin{thebibliography}{14}
\providecommand{\natexlab}[1]{#1}
\providecommand{\url}[1]{\texttt{#1}}
\expandafter\ifx\csname urlstyle\endcsname\relax
  \providecommand{\doi}[1]{doi: #1}\else
  \providecommand{\doi}{doi: \begingroup \urlstyle{rm}\Url}\fi

\bibitem[Breiman(2001)]{breiman2001random}
Leo Breiman.
\newblock Random forests.
\newblock \emph{Machine Learning}, 45\penalty0 (1):\penalty0 5--32, 2001.

\bibitem[Breiman et~al.(1984)Breiman, Friedman, Olshen, and
  Stone]{breiman1984cart}
Leo Breiman, Jerome Friedman, Richard Olshen, and Charles Stone.
\newblock Cart.
\newblock \emph{Classification and regression trees}, 1984.

\bibitem[Chen and Guestrin(2016)]{chen2016xgboost}
Tianqi Chen and Carlos Guestrin.
\newblock Xgboost: A scalable tree boosting system.
\newblock In \emph{Proceedings of the 22nd ACM sigkdd International Conference
  on Knowledge Discovery and Data Mining}, pages 785--794, 2016.

\bibitem[Guyon et~al.(2002)Guyon, Weston, Barnhill, and Vapnik]{guyon2002gene}
Isabelle Guyon, Jason Weston, Stephen Barnhill, and Vladimir Vapnik.
\newblock Gene selection for cancer classification using support vector
  machines.
\newblock \emph{Machine Learning}, 46\penalty0 (1):\penalty0 389--422, 2002.

\bibitem[Harris et~al.(2020)Harris, Millman, van~der Walt, Gommers, Virtanen,
  Cournapeau, Wieser, Taylor, Berg, Smith, Kern, Picus, Hoyer, van Kerkwijk,
  Brett, Haldane, del R{\'{i}}o, Wiebe, Peterson, G{\'{e}}rard-Marchant,
  Sheppard, Reddy, Weckesser, Abbasi, Gohlke, and Oliphant]{harris2020array}
Charles~R. Harris, K.~Jarrod Millman, St{\'{e}}fan~J. van~der Walt, Ralf
  Gommers, Pauli Virtanen, David Cournapeau, Eric Wieser, Julian Taylor,
  Sebastian Berg, Nathaniel~J. Smith, Robert Kern, Matti Picus, Stephan Hoyer,
  Marten~H. van Kerkwijk, Matthew Brett, Allan Haldane, Jaime~Fern{\'{a}}ndez
  del R{\'{i}}o, Mark Wiebe, Pearu Peterson, Pierre G{\'{e}}rard-Marchant,
  Kevin Sheppard, Tyler Reddy, Warren Weckesser, Hameer Abbasi, Christoph
  Gohlke, and Travis~E. Oliphant.
\newblock Array programming with {NumPy}.
\newblock \emph{Nature}, 585\penalty0 (7825):\penalty0 357--362, September
  2020.
\newblock \doi{10.1038/s41586-020-2649-2}.
\newblock URL \url{https://doi.org/10.1038/s41586-020-2649-2}.

\bibitem[Kaggle(2021)]{lendingclub2021}
Kaggle.
\newblock Lending club loan data.
\newblock
  \url{https://www.kaggle.com/datasets/adarshsng/lending-club-loan-data-csv},
  Jun 2021.

\bibitem[Liu et~al.(2022)Liu, Hu, Chen, and Nair]{liu2022performance}
Alice~J. Liu, Linwei Hu, Jie Chen, and Vijayan Nair.
\newblock Performance and interpretability comparisons of supervised machine
  learning algorithms: An empirical study.
\newblock \emph{arXiv preprint arXiv:2204.12868}, 2022.

\bibitem[Lloyd(1982)]{lloyd1982least}
Stuart Lloyd.
\newblock Least squares quantization in pcm.
\newblock \emph{IEEE transactions on information theory}, 28\penalty0
  (2):\penalty0 129--137, 1982.

\bibitem[Nori et~al.(2019)Nori, Jenkins, Koch, and
  Caruana]{nori2019interpretml}
Harsha Nori, Samuel Jenkins, Paul Koch, and Rich Caruana.
\newblock Interpretml: A unified framework for machine learning
  interpretability.
\newblock \emph{arXiv preprint arXiv:1909.09223}, 2019.

\bibitem[Potdar et~al.(2017)Potdar, Pardawala, and Pai]{potdar2017comparative}
Kedar Potdar, Taher~S. Pardawala, and Chinmay~D. Pai.
\newblock A comparative study of categorical variable encoding techniques for
  neural network classifiers.
\newblock \emph{International Journal of Computer Applications}, 175\penalty0
  (4):\penalty0 7--9, 2017.

\bibitem[Snoek et~al.(2012)Snoek, Larochelle, and Adams]{snoek2012practical}
Jasper Snoek, Hugo Larochelle, and Ryan~P. Adams.
\newblock Practical bayesian optimization of machine learning algorithms.
\newblock \emph{Advances in neural information processing systems}, 25, 2012.

\bibitem[Sudjianto and Zhang(2021)]{sudjianto2021designing}
Agus Sudjianto and Aijun Zhang.
\newblock Designing inherently interpretable machine learning models.
\newblock \emph{arXiv preprint arXiv:2111.01743}, 2021.

\bibitem[Sundstr{\"o}m(2010)]{sundstrom2010coding}
Stina Sundstr{\"o}m.
\newblock Coding in multiple regression analysis: A review of popular coding
  techniques.
\newblock 2010.

\bibitem[Tibshirani(1996)]{tibshirani1996regression}
Robert Tibshirani.
\newblock Regression shrinkage and selection via the lasso.
\newblock \emph{Journal of the Royal Statistical Society: Series B
  (Methodological)}, 58\penalty0 (1):\penalty0 267--288, 1996.

\end{thebibliography}
